\documentclass[traditabstract,longauth]{aa}
\usepackage{graphicx}
\usepackage{txfonts}
\usepackage[colorlinks=true,citecolor=blue,backref=page,breaklinks=true]{hyperref}
\usepackage{xspace}
\usepackage{amsmath}
\usepackage{tabularx}
\usepackage{booktabs}
\usepackage{scalerel}
\usepackage{longtable}
\usepackage{threeparttable}
\usepackage{lineno}

\usepackage{natbib,twoopt}
\bibpunct{(}{)}{;}{a}{}{,}             
\makeatletter
  \newcommandtwoopt{\citeads}[3][][]{\href{http://adsabs.harvard.edu/abs/#3}%
    {\def\hyper@linkstart##1##2{}%
     \let\hyper@linkend\@empty\citealp[#1][#2]{#3}}}
  \newcommandtwoopt{\citepads}[3][][]{\href{http://adsabs.harvard.edu/abs/#3}%
    {\def\hyper@linkstart##1##2{}%
     \let\hyper@linkend\@empty\citep[#1][#2]{#3}}}
  \newcommandtwoopt{\citetads}[3][][]{\href{http://adsabs.harvard.edu/abs/#3}%
    {\def\hyper@linkstart##1##2{}%
     \let\hyper@linkend\@empty\citet[#1][#2]{#3}}}
  \newcommandtwoopt{\citeyearads}[3][][]%
    {\href{http://adsabs.harvard.edu/abs/#3}
    {\def\hyper@linkstart##1##2{}%
     \let\hyper@linkend\@empty\citeyear[#1][#2]{#3}}}
\makeatother

\usepackage{xcolor}
\newcommand{\orcidicon}[1]{}
\usepackage{hyperref}

\usepackage[normalem]{ulem}

\newcommand{\rev}[1]{{#1}}

\newcommand{\tess}{\textit{TESS}\xspace}
\newcommand{\gaia}{\textit{Gaia}\xspace}

\newcommand{\pyttv}{\texttt{PyTTV}\xspace}
\newcommand{\pytransit}{\texttt{PyTransit}\xspace}
\newcommand{\celerite}{\texttt{celerite}\xspace}
\newcommand{\emcee}{\texttt{emcee}\xspace}
\newcommand{\rebound}{\texttt{Rebound}\xspace}
\newcommand{\reboundx}{\texttt{Reboundx}\xspace}
\newcommand{\vosa}{\texttt{VOSA}\xspace}
\newcommand{\ispec}{\texttt{iSpec}\xspace}
\newcommand{\param}{\texttt{PARAM\,1.5}\xspace}
\newcommand{\sme}{\texttt{SME}\xspace}
\newcommand{\TTF}{\texttt{\tess Transit Finder}\xspace}
\newcommand{\tapir}{\texttt{Tapir}\xspace}
\newcommand{\banzai}{\texttt{BANZAI}\xspace}
\newcommand{\astroi}{\texttt{AstroImageJ}\xspace}
\newcommand{\munipack}{\texttt{C-Munipack}\xspace}
\newcommand{\exostriker}{\texttt{EXOSTRIKER}\xspace}

\newcommand{\p}{\ensuremath{P}\xspace}
\newcommand{\epoch}{\ensuremath{T_0}\xspace}
\newcommand{\tc}{\ensuremath{T_\mathrm{c}}\xspace}
\newcommand{\rpoverrstar}{\ensuremath{R_\mathrm{p}/R_\star}\xspace}
\newcommand{\pimp}{\ensuremath{b}\xspace}
\newcommand{\qone}{\ensuremath{q_1}\xspace}
\newcommand{\qtwo}{\ensuremath{q_2}\xspace}
\newcommand{\rhostar}{\ensuremath{\rho_\star}\xspace}
\newcommand{\kp}{\ensuremath{k}\xspace}

\newcommand{\ecc}{\ensuremath{e}\xspace}
\newcommand{\omegaorbit}{\ensuremath{\omega}\xspace}
\newcommand{\Omegaorbit}{\ensuremath{\Omega}\xspace}

\newcommand{\UP}[1]{\ensuremath{\mathcal{U}(#1)}\xspace}
\newcommand{\NP}[1]{\ensuremath{\mathcal{N}(#1)}\xspace}

\newcommand{\mjup}{\ensuremath{M_\mathrm{Jup}}\xspace}
\newcommand{\rjup}{\ensuremath{R_\mathrm{Jup}}\xspace}
\newcommand{\mearth}{\ensuremath{M_\oplus}\xspace}
\newcommand{\rearth}{\ensuremath{R_\oplus}\xspace}
\newcommand{\smass}{\ensuremath{M_\star}\xspace}
\newcommand{\sradius}{\ensuremath{R_\star}\xspace}
\newcommand{\msun}{\ensuremath{M_\odot}\xspace}
\newcommand{\rsun}{\ensuremath{R_\odot}\xspace}
\newcommand{\teff}{\ensuremath{T_{\mathrm{eff}}}\xspace}
\newcommand{\logg}{\ensuremath{\log g}\xspace}
\newcommand{\vsini}{\ensuremath{v_\mathrm{rot}\sin{i_\star}}\xspace}
\newcommand{\lum}{\ensuremath{L_\star}\xspace}

\newcommand{\feh}{[Fe/H]\xspace}
\newcommand{\Av}{\ensuremath{A_\mathrm{v}}\xspace}

\newcommand{\msununit}{\ensuremath{M_\odot}}
\newcommand{\rsununit}{\ensuremath{R_\odot}}
\newcommand{\fehunit}{[Fe/H]}

\newcommand{\kms}{km\,$\mathrm{s}^{-1}$}
\newcommand{\msunit}{m\,$\mathrm{s}^{-1}$}

\begin{document} 


\title{TOI-1130: A photodynamical analysis of a hot Jupiter in resonance with an inner low-mass planet \thanks{Based on observations made with ESO 3.6-m telescope at La Silla Observatory under programme IDs 1102.C-0923 and 60.A-9709. This paper includes data gathered with the 6.5 meter Magellan Telescopes located at Las Campanas Observatory, Chile.}}
\author{J.~Korth\inst{\ref{Chalmers},\ref{lund}}\fnmsep\orcidicon{0000-0002-0076-6239}
        \and D.~Gandolfi\inst{\ref{TUR}}\orcidicon{0000-0001-8627-9628} 
        \and J.~\v{S}ubjak\inst{\ref{Ondr},\ref{UfP},\ref{ESO1},\ref{CfA}}\orcidicon{0000-0002-5313-9722}
        \and S.~Howard\inst{\ref{CNRS}}\orcidicon{0000-0003-4894-7271}
        \and S.~Ataiee\inst{\ref{FUM}}\orcidicon{0000-0003-2594-3454}
        \and   K.~A.~Collins\inst{\ref{CfA}}\orcidicon{0000-0001-6588-9574}
        \and S.~N.~Quinn\inst{\ref{CfA}}\orcidicon{0000-0002-8964-8377}
        \and A.~J.~Mustill\inst{\ref{lund}}\orcidicon{0000-0002-2086-3642}
        \and
        T.~Guillot\inst{\ref{CNRS}}\orcidicon{0000-0002-7188-8428}
        \and 
        N.~Lodieu\inst{\ref{IAC}, \ref{ULL}}\orcidicon{0000-0002-3612-8968}
        \and A.~M.~S.~Smith\inst{\ref{DLR}}\orcidicon{0000-0002-2386-4341}
        \and M.~Esposito\inst{\ref{TLS}}\orcidicon{0000-0002-6893-4534}
        \and F.~Rodler\inst{\ref{ESO}}\orcidicon{0000-0003-0650-5723} 
        \and A.~Muresan\inst{\ref{Chalmers}}\orcidicon{0000-0002-3439-4330}
        \and
        L.~Abe\inst{\ref{CNRS}}\orcidicon{0000-0002-0856-4527}
        \and S.~H.~Albrecht\inst{\ref{Aar}}\orcidicon{0000-0003-1762-8235}
        \and 
        A.~Alqasim\inst{\ref{MUL}}
        \and
        K.~Barkaoui\inst{\ref{liege},\ref{MIT},\ref{IAC}}\orcidicon{0000-0003-1464-9276}
        \and P.~G.~Beck\inst{\ref{GRAZ},\ref{IAC}}\orcidicon{0000-0003-4745-2242}
        \and C.~J.~Burke\inst{\ref{kavali}}\orcidicon{0000-0002-7754-9486} 
        \and R.~P.~Butler\inst{\ref{carnegie}}\orcidicon{0000-0003-1305-3761}
        \and D.~M.~Conti\inst{\ref{AAVS}}\orcidicon{0000-0003-2239-0567}
        \and K.~I.~Collins\inst{\ref{fair}}\orcidicon{0000-0003-2781-3207} 
        \and J.~D.~Crane\inst{\ref{carnegieobs}}\orcidicon{0000-0002-5226-787X}
        \and F.~Dai\inst{\ref{fei}}\orcidicon{0000-0002-8958-0683} 
        \and H.~J.~Deeg\inst{\ref{IAC},\ref{ULL}}\orcidicon{0000-0003-0047-4241}
        \and P.~Evans\inst{\ref{sauce}}\orcidicon{0000-0002-5674-2404}
        \and S.~Grziwa\inst{\ref{RIU}}\orcidicon{0000-0003-3370-4058}
        \and A.~P.~Hatzes\inst{\ref{TLS}}
        \and T.~Hirano\inst{\ref{astrojapan}, \ref{obsjapan}}\orcidicon{0000-0003-3618-7535} 
        \and K.~Horne\inst{\ref{andrews}}\orcidicon{0000-0003-1728-0304}
        \and C.~X.~Huang\inst{\ref{USQ}}\orcidicon{0000-0003-0918-7484} 
        \and J.~M.~Jenkins\inst{\ref{nasa}}\orcidicon{0000-0002-4715-9460}
        \and P.~Kab\'{a}th\inst{\ref{Ondr}}\orcidicon{0000-0002-1623-5352}
        \and J.~F.~Kielkopf\inst{\ref{Loui}}\orcidicon{0000-0003-0497-2651}
        \and E.~Knudstrup\inst{\ref{Aar}}\orcidicon{0000-0001-7880-594X}
        \and D.~W.~Latham\inst{\ref{CfA}}\orcidicon{0000-0001-9911-7388}
        \and J.~Livingston\inst{\ref{astrojapan}, \ref{obsjapan}, \ref{sokendai}}
        \and R.~Luque\inst{\ref{UC}}\orcidicon{0000-0002-4671-2957}
        \and S.~Mathur\inst{\ref{IAC},\ref{ULL}}\orcidicon{0000-0002-0129-0316} 
        \and F.~Murgas\inst{\ref{IAC},\ref{ULL}}\orcidicon{0000-0001-9087-1245}
        \and H.~L.~M.~Osborne\inst{\ref{MUL}}\orcidicon{0000-0002-4143-4767}
        \and E.~Palle\inst{\ref{IAC},\ref{ULL}}
        \and C~M.~Persson\inst{\ref{OSO}} 
        \and J.~E.~Rodriguez\inst{\ref{michigan}}\orcidicon{0000-0001-8812-0565}
        \and M.~Rose\inst{\ref{nasa}}\orcidicon{0000-0003-4724-745X} 
        \and P.~Rowden\inst{\ref{RAS}}\orcidicon{0000-0002-4829-7101} 
        \and R.~P.~Schwarz\inst{ \ref{CfA}}\orcidicon{0000-0001-8227-1020}
        \and S.~Seager\inst{\ref{kavali},\ref{MIT},\ref{MIT_aero}}\orcidicon{0000-0002-6892-6948} 
        \and L.~M.~Serrano\inst{\ref{TUR}}\orcidicon{0000-0001-9211-3691}
        \and L.~Sha\inst{\ref{wisconsin}}\orcidicon{0000-0001-5401-8079}
        \and S.~A.~Shectman\inst{\ref{carnegieobs}}\orcidicon{0000-0002-8681-6136}
        \and A.~Shporer\inst{\ref{kavali}}\orcidicon{0000-0002-1836-3120}
        \and G.~Srdoc\inst{\ref{Koti}} 
        \and C.~Stockdale\inst{\ref{hazelwood}}\orcidicon{0000-0003-2163-1437}
        \and T.-G.~Tan\inst{\ref{pest}}\orcidicon{0000-0001-5603-6895}
        \and J.~K.~Teske\inst{\ref{carnegie}}
        \and V.~Van Eylen\inst{\ref{MUL}}\orcidicon{0000-0001-5542-8870}
        \and A.~Vanderburg\inst{\ref{kavali}}\orcidicon{0000-0001-7246-5438}
        \and R.~Vanderspek\inst{\ref{kavali}}\orcidicon{0000-0001-6763-6562}
        \and S.~X.~Wang\inst{\ref{tsinghua}}\orcidicon{0000-0002-6937-9034}
        \and J.~N.~Winn\inst{\ref{princeton}}\orcidicon{0000-0002-4265-047X}
          }
        \offprints{judith.korth@fysik.lu.se}
   \institute{
        Department of Space, Earth and Environment, Astronomy and Plasma Physics, Chalmers University of Technology, Chalmersplatsen 4, 412 96 Gothenburg, Sweden\label{Chalmers}
        \and \rev{Lund Observatory, Division of Astrophysics, Department of Physics, Lund University, Box 43, 22100 Lund, Sweden}\label{lund}
        \and Dipartimento di Fisica, Universit\`a degli Studi di Torino, via Pietro Giuria 1, I-10125, Torino, Italy\label{TUR}
        \and Astronomical Institute of the Czech Academy of Sciences, Fri\v{c}ova 298, 25165, Ond\v{r}ejov, Czech Republic\label{Ondr}
        \and Astronomical Institute of Charles University, V Hole\v{s}ovi\v{c}k\'ach 2, 180 00, Prague, Czech Republic\label{UfP}
        \and ESO, Karl-Schwarzschild-Stra{\ss}e 2, 85748 Garching bei M\"unchen, Germany\label{ESO1}
        \and Center for Astrophysics \textbar \ Harvard \& Smithsonian, 60 Garden Street, Cambridge, MA 02138, USA\label{CfA}
        \and Universit\'e C\^ote d'Azur, Observatoire de la C\^ote d'Azur, CNRS, Laboratoire Lagrange, Bd de l'Observatoire, CS 34229, 06304 Nice cedex 4, France \label{CNRS}
        \and Department of Physics, Faculty of Sciences, Ferdowsi University of Mashhad, Mashhad, 91775-1436, Iran\label{FUM}
        \and Instituto de Astrof\'{\i}sica de Canarias (IAC), E-38200 La Laguna, Tenerife, Spain\label{IAC}
        \and Universidad de La Laguna (ULL), Departamento de Astrof\'\i sica, E-38206 La Laguna, Tenerife, Spain\label{ULL}
        \and Institute for Planetary Research, German Aerospace Center (DLR), Rutherfordstr. 2, 12489 Berlin, Germany\label{DLR}
        \and Thüringer Landessternwarte Tautenburg, Sternwarte 5, D-07778 Tautenburg, Germany\label{TLS}
        \and European Southern Observatory, Alonso de Cordova 3107, Vitacura, Santiago de Chile, Chile\label{ESO}
        \and Stellar Astrophysics Centre, Department of Physics and Astronomy, Aarhus University, Ny Munkegade 120, DK-8000 Aarhus C, Denmark\label{Aar}
        \and Mullard Space Science Laboratory, University College London, Holmbury St Mary, Dorking, Surrey RH5 6NT, UK\label{MUL}
        \and Astrobiology Research Unit, Universit\'e de Li\`ege, All\'ee du 6 Ao\^ut 19C, B-4000 Li\`ege, Belgium\label{liege}
        \and Department of Earth, Atmospheric and Planetary Science, Massachusetts Institute of Technology, 77 Massachusetts Avenue, Cambridge, MA 02139, USA\label{MIT}
        \and Institut für Physik, Karl-Franzens Universität Graz, Universitätsplatz 5/II, NAWI Graz, 8010 Graz, Austria\label{GRAZ}
        \and Department of Physics and Kavli Institute for Astrophysics and Space Research, Massachusetts Institute of Technology, Cambridge, MA 02139, USA\label{kavali}
        \and Carnegie Earth and Planets Laboratory, 5241 Broad Branch Road NW, Washington, DC 20015, USA\label{carnegie}
        \and American Association of Variable Star Observers, 49 Bay State Road, Cambridge, MA 02138, USA\label{AAVS}
        \and George Mason University, 4400 University Drive, Fairfax, VA, 22030 USA\label{fair}
        \and The Observatories of the Carnegie Institution for Science, 813 Santa Barbara St., Pasadena, CA 91101, USA\label{carnegieobs}
        \and Division of Geological and Planetary Sciences, 1200 E California Blvd, Pasadena, CA, 91125, USA\label{fei}
        \and El Sauce Observatory, Coquimbo Province, Chile\label{sauce}
        \and Rheinisches Institut für Umweltforschung, Abt. Planetenforschung, an der Universität zu Köln, 50931 Cologne, Germany\label{RIU}
        \and Astrobiology Center, 2-21-1 Osawa, Mitaka, Tokyo 181-8588, Japan\label{astrojapan}
        \and National Astronomical Observatory of Japan, 2-21-1 Osawa, Mitaka, Tokyo 181-8588, Japan\label{obsjapan}
        \and SUPA School of Physics \& Astronomy, North Haugh, University of St~Andrews, St~Andrews KY16~9SS, Scotland, UK\label{andrews}
        \and University of Southern Queensland, Centre for Astrophysics, West Street, Toowoomba, QLD 4350 Australia\label{USQ}
        \and NASA Ames Research Center, Moffett Field, CA 94035, USA\label{nasa}
        \and Department of Physics and Astronomy, University of Louisville, Louisville, KY 40292, USA\label{Loui}
        \and Department of Astronomy, The Graduate University for Advanced Studies (SOKENDAI), 2-21-1 Osawa, Mitaka, Tokyo, Japan\label{sokendai}
        \and Department of Astronomy \& Astrophysics, University of Chicago, Chicago, IL 60637, USA\label{UC}
        \newpage
        \and Department of Space, Earth and Environment, Chalmers University of Technology, Onsala Space Observatory, SE-439 92 Onsala, Sweden. \label{OSO} 
        \and Center for Data Intensive and Time Domain Astronomy, Department of Physics and Astronomy, Michigan State University, East Lansing, MI 48824, USA\label{michigan}
        \and Royal Astronomical Society, Burlington House, Piccadilly, London W1J 0BQ\label{RAS}
        \and Patashnick Voorheesville Observatory, Voorheesville, NY 12186, USA\label{Voor}
        \and Department of Aeronautics and Astronautics, MIT, 77 Massachusetts Avenue, Cambridge, MA 02139, USA\label{MIT_aero}
        \and Department of Astronomy, University of Wisconsin--Madison, 475 N Charter St, Madison, WI 53706, USA\label{wisconsin}
        \and Kotizarovci Observatory, Sarsoni 90, 51216 Viskovo, Croatia\label{Koti}
        \and Hazelwood Observatory, Australia\label{hazelwood}
        \and Perth Exoplanet Survey Telescope, Perth, Western Australia\label{pest}
        \and Department of Astronomy, Tsinghua University, Beijing 100084, People's Republic of China\label{tsinghua}
        \and Department of Astrophysical Sciences, Princeton University, Princeton, NJ 08544, USA\\\newpage \label{princeton}
             }
   \date{Received September 15, 1996; accepted March 16, 1997}

 
  \abstract
     {The TOI-1130 is a known planetary system around a K-dwarf consisting of a gas giant planet, TOI-1130\,c, on an 8.4-day orbit, accompanied by an inner Neptune-sized planet, TOI-1130\,b, with an orbital period of 4.1 days. \rev{We collected precise radial velocity (RV) measurements of TOI-1130 with the HARPS and PFS spectrographs as} part of our ongoing RV follow-up program. We perform a photodynamical modeling of the HARPS and PFS RVs, and transit photometry from the Transiting Exoplanet Survey Satellite (\tess) and the \tess Follow-up Observing Program. We determine the planet masses and radii of TOI-1130\,b and TOI-1130\,c to be \rev{$M_\mathrm{b}\,=19.28\pm\,0.97$\,\mearth and $R_\mathrm{b}\,=3.56\pm\,0.13$\,\rearth, and $M_\mathrm{c}\,=325.59\pm\,5.59$\,\mearth and $R_\mathrm{c}\,=13.32^{+1.55}_{-1.41}$\,\rearth, respectively}. We spectroscopically confirm TOI-1130\,b that was previously only validated. We find that the two planets orbit with small eccentricities in a 2:1 resonant configuration. This is the first known system with a hot Jupiter and an inner lower mass planet locked in a mean-motion resonance. TOI-1130 belongs to the small yet increasing population of hot Jupiters with an inner low-mass planet that challenges the pathway for hot Jupiter formation. We also detect a linear RV trend \rev{possibly} due to the presence of an outer massive companion.}


   \maketitle
%

\section{Introduction}

The diversity within the exoplanet ``jungle'' is one of the astonishing outcomes in exoplanet research over the last 30 years. Exoplanets are found to survive in hostile environments in orbits with very short orbital periods. The first exoplanet detected around the solar-like star 51\,Peg \citepads{1995Natur.378..355M}, represents one of these new types of planets, a gas giant on a short-period orbit ($P_\mathrm{orb}<10$ day), also known as a hot Jupiter. Even 27 years after this discovery, there is no clear picture of their formation, whether they formed in situ or further out beyond the ice line and migrated inwards (see \citetads{2021JGRE..12606629F} and references therein). 

The in situ formation mechanism is proposed to happen at the present-day close-in orbit when a core accretes gas from the gaseous protoplanetary disks (\citeads{2016ApJ...817L..17B}; \citeads{2018ApJ...866L...2B}). If sufficient material is available close to the star to build up a $\sim$\,10\,\mearth core \citepads{2006ApJ...648..666R}\rev{, and the other conditions, such as planetesimal accretion luminosity and gas opacity, are right \citepads{2014ApJ...797...95L}}, the core will accrete gas as long as the gaseous protoplanetary disk is not dissipated, and form a gas giant \citepads{2018ARA&A..56..175D}. 
In the migration theory, it is assumed that all gas giants form beyond the ice-line \citepads{2009Icar..200..672D}, and some of them migrate close to the host star to become hot Jupiters \citepads{2018ARA&A..56..175D}. This migration is thought to happen either through the interaction with the gas disk during the formation period (\citeads{1996Natur.380..606L}; \citeads{2004MNRAS.350..849N}; \citeads{2012ARA&A..50..211K}; \citeads{2019A&A...623A..88B}), or via high-eccentricity migration (HEM; \citeads{1996Sci...274..954R}; \citeads{2015ApJ...808...14M}) at a later stage. In the HEM scenario the eccentricities are either excited by planet--planet scattering (\citeads{1996Sci...274..954R}; \citeads{2008ApJ...686..580C}), by the Kozai--Lidov cycles \citepads{2003ApJ...589..605W}, or by interactions with a companion (\citeads{2011ApJ...735..109W}; \citeads{2015ApJ...805...75P}). Recent discoveries suggest that HEM is the dominant mechanism \rev{(\citeads{2019MNRAS.484.5645V}, \citeyearads{2023ApJ...943L..13V}; \citeads{2023AJ....165...82J})}. 

Hot Jupiters are often accompanied by gas giants on wider orbits (\citeads{2014ApJ...785..126K}) but rarely accompanied by \rev{aligned} nearby planets (\citeads{2012PNAS..109.7982S}; \citeads{2016ApJ...825...98H}; \citeads{2021AJ....162..263H}; \citeads{2022ApJS..259...62I}). 
The absence of low-mass planets in systems containing a hot Jupiter is one of the key arguments in support of HEM over the in situ formation. 

However, systems have been detected in which a hot gas giant is accompanied by an inner low-mass planet. The first example is WASP-47 (\citeads{2012MNRAS.426..739H}; \citeads{2015ApJ...812L..18B}; \citeads{2022AJ....163..197B}; \citeads{2023A&A...673A..42N}). Since then, more systems have been detected: Kepler-730 (\citeads{2018RNAAS...2..160Z}; \citeads{2019ApJ...870L..17C}), TOI-2000 \citepads{2022arXiv220914396S}, WASP-132  (\citeads{2017MNRAS.465.3693H}; \citeads{2022AJ....164...13H}), and TOI-1130 \citepads{2020ApJ...892L...7H}, that we discuss here in more detail. \rev{Moreover, there is some indication from transit timing variation
(TTV; \citeads{2005MNRAS.359..567A}) measurements that non-aligned nearby companions to hot
Jupiters may be more common than previously thought, although
more work is needed to expand the sample size \citepads{2023AJ....165..171W}.} Those systems rule out HEM, which allows the formation of outer companions and prohibits the formation of inner companions. Instead, those systems could have formed through disk migration (\citeads{2003ApJ...599L.111M}; \citeads{2005A&A...441..791F}, \citeyearads{2007A&A...472.1003F}; \citeads{2014ApJ...787..172O}) or in situ \citepads{2021MNRAS.505.2500P}, since both scenarios allow the formation of terrestrial planets inside the orbit of a hot Jupiter. \rev{Compared to the other systems that contain a hot Jupiter and an inner low-mass planet, TOI-1130 shows a unique orbital configuration: their orbital periods are close to a 2:1 period commensurability indicating that this system could be in a first-order mean motion resonance (MMR). If confirmed TOI-1130 would have formed most likely via disk migration (\citeads{2011MNRAS.413..554M}; \citeads{2018CeMDA.130...54P}) and not via in situ formation.} Thus, we can distinguish between different formation scenarios by knowing the system's architecture. In situ formation permits the formation of nearby planets, disk migration permits the formation of nearby planets in resonant orbits, and HEM permits the formation of outer planets. 

In this article, we present a study of the architecture of TOI-1130, a system that contains a gas giant (TOI-1130\,c) and a lower mass planet (TOI-1130\,b) detected by \citetads{2020ApJ...892L...7H}. We carried out spectroscopic ground-based follow-up of TOI-1130 to determine the planetary and orbital parameters, especially for the inner planet TOI-1130\,b, which had previously only been validated. \rev{Since the orbital period ratio of the two planets is close to a 2:1 period commensurability, we expect to measure large TTVs as already reported in \citetads{2020ApJ...892L...7H}.} Thus, we acquired photometric follow-up to have a good phase coverage of the expected TTV signal. The ground-based photometry is modeled photodynamically together with the Transiting Exoplanet Survey Satellite (\tess) and radial velocity (RV) data to determine precisely the orbital and planetary parameters.

\section{Observation and data reduction}

We here provide a brief description of the TOI-1130 observations and time-series data used in the subsequent analysis: the space-based \tess\ photometry (Sect.~\ref{TESS_photometry}), the ground-based photometry (Sect.~\ref{Ground_based_photometry}), and the high-resolution spectroscopy (Sect.~\ref{HARPS_PFS_spectroscopy}). 

\subsection{\tess photometry}
\label{TESS_photometry}

TOI-1130 was observed by \tess in Sectors~13 and 27 in the southern ecliptic hemisphere. Sector~13 was observed between 2019 Jul 18 and 19, covering six transits of TOI-1130\,b and three transits of TOI-1130\,c. Sector 27 was observed between 2020 Jul 5 and 30 spanning six transits of TOI-1130\,b and three transits of TOI-1130\,c. While TOI-1130 was observed in the 30-min cadence mode in Sector~13 (Camera 2, CCD 1), it was observed in Sector~27 with a higher cadence rate\rev{s} of \rev{2-min and} 20~s (Camera 1, CCD 1).  

We used in the subsequent analysis the publicly available $\mathrm{KSPSAP\_FLUX}$ light curves produced by the MIT Quick-Look-Pipeline (QLP; \citeads{2020RNAAS...4..204H}, \citeyearads{2020RNAAS...4..206H}; \citeads{2021RNAAS...5..234K}) and the Presearch Data Conditioning (PDC) light curves (\citeads{2012PASP..124.1000S}; \citeads{2014PASP..126..100S}) produced by the Science Processing Operations Center (SPOC; \citeads{2016SPIE.9913E..3EJ}) \rev{at NASA Ames Research Center} downloaded from the Mikulski Archive for Space Telescopes\footnote{\url{https://mast.stsci.edu}.} for Sector~13 and Sector~27, respectively. We note that \citetads{2020ApJ...892L...7H} based their study on their own photometry created for Sector~13. 

\subsection{Ground-based photometry}
\label{Ground_based_photometry}

\begin{table*}[!ht]
\begin{threeparttable}
 \centering
 \small
 \caption{Ground-based light curve observations of TOI-1130\,b and TOI-1130\,c.}
 \label{table:ground-obs-table}
	\begin{tabularx}{\textwidth}{@{\extracolsep{\fill}}lclcccc}
    \toprule
    \toprule
Observatory  & Aperture [m] & Location                 & UTC Date      & Filter                  & Planet & Transit number\\
    \midrule
LCOGT-SSO   & 1.0  & Siding Spring, Australia  & 2019-09-05   &  Pan-STARRS $z$-short   & b & 19\tablefootmark{a} \\
PEST       & 0.31 & Perth, Australia          & 2019-10-01   &  $R_c$              & c & 13\tablefootmark{a} \\
LCOGT-SSO          &  1.0 &  Siding Spring, Australia  & 2020-05-05   &  Pan-STARRS $z$-short  & c & 39\\
LCOGT-SSO          &  1.0 &  Siding Spring, Australia  & 2020-05-05   &  $B$  & c & 39\\
LCOGT-SAAO             & 1.0  & Sutherland, South Africa   & 2020-06-07   &  Pan-STARRS $z$-short   & c & 43 \\
LCOGT-SAAO             & 1.0  & Sutherland, South Africa   & 2020-06-07   &  Pan-STARRS $z$-short   & c & 43 \\
LCOGT-SSO            & 1.0  & Siding Spring, Australia  & 2020-08-05   &  Pan-STARRS $z$-short   & c & 50 \\
El Sauce Observatory & 0.36 & Coquimbo Province, Chile  & 2020-08-22   &  $R_c$              & c & 52 \\
PEST                 & 0.31 & Perth, Australia          & 2020-08-30   &  $R_c$              & c & 53 \\
LCOGT-SAAO             & 1.0  & Sutherland, South Africa   & 2021-04-12   &  Pan-STARRS $z$-short   & c & 80 \\
LCOGT-SAAO             & 1.0  & Sutherland, South Africa   & 2021-06-10   &  Sloan $i'$             & b & 177 \\
LCOGT-CTIO             & 1.0  & Cerro Tololo, Chile       & 2021-06-19   &  Sloan $i'$             & b & 179 \\
LCOGT-CTIO             & 1.0  & Cerro Tololo, Chile       & 2021-06-26   &  Sloan $i'$             & c & 88 \\
LCOGT-CTIO             & 1.0  & Cerro Tololo, Chile       & 2021-06-27   &  Sloan $i'$             & c & 88 \\
LCOGT-SSO            & 1.0  & Siding Spring, Australia  & 2021-07-30   &  Sloan $i'$             & c & 93\\
LCOGT-SAAO             & 1.0  & Sutherland, South Africa   & 2021-08-02   &  Sloan $i'$             & b & 190\\
LCOGT-CTIO             & 1.0  & Cerro Tololo, Chile       & 2021-08-06   &  Sloan $i'$             & b & 191 \\
LCOGT-SAAO             & 1.0  & Sutherland, South Africa   & 2021-08-07   &  Sloan $i'$             & c & 94\\
LCOGT-CTIO             & 1.0  & Cerro Tololo, Chile       & 2021-08-10   &  Sloan $i'$             & b & 192 \\
ASTEP       & 0.4  & East Antarctic plateau    & 2021-08-24   &  Similar to $R_c$   & c & 96 \\
    \bottomrule
    \bottomrule
    \end{tabularx}
  \end{threeparttable}
  \tablefoot{
\tablefoottext{a}{Published in \citetads{2020ApJ...892L...7H}.}
}
\end{table*}
We acquired ground-based time-series photometry of TOI-1130 as part of the \tess Follow-up Observing Program (TFOP; \citeads{2018AAS...23143908C}; \citeads{2019AAS...23314005C})\footnote{\url{https://tess.mit.edu/followup}.}. In addition to the photometry published in \citetads{2020ApJ...892L...7H}, we observed five transits of TOI-1130\,b and 13 of TOI-1130\,c as listed in Table~\ref{table:ground-obs-table}. We used the \TTF, which is a customized version of the \tapir software package \citepads{2013ascl.soft06007J}, to schedule our transit observations.

We observed 16 transits using the Las Cumbres Observatory Global Telescope (LCOGT; \citeads{2013PASP..125.1031B}) 1.0-m network. The telescopes are equipped with $4096\times4096$ Sinistro cameras having an image scale of $0\farcs389$ per pixel, resulting in a $26\arcmin\times26\arcmin$ field of view. The images were calibrated by the standard LCOGT \banzai pipeline \citepads{2018SPIE10707E..0KM}, and photometric data were extracted using \astroi \citepads{2017AJ....153...77C}.

We observed two transits from the Perth Exoplanet Survey Telescope (PEST) near Perth, Australia. The 0.3-m telescope is equipped with a $1530\times1020$ SBIG ST-8XME camera with an image scale of 1$\farcs$2 pixel$^{-1}$, resulting in a $31\arcmin\times21\arcmin$ field of view. A custom pipeline based on \munipack\footnote{\url{http://c-munipack.sourceforge.net}.} was used to calibrate the images and extract the differential photometry.

We observed one transit from the Evans 0.36-m telescope at El Sauce Observatory in Coquimbo Province, Chile. The telescope is equipped with a $1536\times1024$ SBIG STT-1603-3 camera. The image scale is 1$\farcs$47 pixel$^{-1}$ \rev{with in-camera binning $2\times2$}, resulting in an $18.8\arcmin\times12.5\arcmin$ field of view. The images were calibrated and photometric data were extracted using \astroi.

The Antarctica Search for Transiting ExoPlanets (ASTEP) program on the East Antarctic plateau (\citeads{2015AN....336..638G}; \citeads{2016MNRAS.463...45M}) also observed one transit. The 0.4-m telescope is equipped with an FLI Proline science camera with a KAF-16801E, $4096\times 4096$ front-illuminated CCD. The camera has an image scale of $0\farcs93$\,pixel$^{-1}$ resulting in a $1^\circ\times1^\circ$ corrected field of view. The data were processed using an automated IDL-based pipeline described in \citetads{2013A&A...553A..49A}.



\subsection{High-resolution spectroscopy}
\label{HARPS_PFS_spectroscopy}

We acquired 49 high-resolution ($R$\,$\approx$\,115\,000) spectra using the High Accuracy Radial velocity Planet Searcher (HARPS; \citeads{2003Msngr.114...20M}) spectrograph mounted at the 3.6-m telescope of the European Southern Observatory (ESO), La Silla, Chile. The observations were performed between 2019 Sep 18 and 19 as part of ESO programs 1102.C-0923 (PI: Gandolfi), and 60.A-9709 (technical night). In total, we observed the target on 41 individual nights with exposure times of 35 minutes. We reduced the data using the HARPS data reduction software (DRS; \citeads{2007A&A...468.1115L}) and extracted the radial velocity by cross-correlating the HARPS spectra with a K5 numerical mask (\citeads{1996A&AS..119..373B}; \citeads{2002A&A...388..632P}) and achieved a mean precision of 1.1\,\msunit. We also used the DRS to measure the Ca {\sc ii} H \& K lines and to calculate the S-index, and extracted the full width at half maximum (FWHM) and the bisector inverse slope (BIS) of the cross-correlation function (CCF). These values are listed in
Table~\ref{tab:toi-1130_harps}. 

We also observed TOI-1130 with the Planet Finder Spectrograph (PFS; \citeads{2006SPIE.6269E..31C}, \citeyearads{2008SPIE.7014E..79C}, \citeyearads{2010SPIE.7735E..53C}), which is mounted on the 6.5-m Magellan II (Clay) Telescope at Las Campanas Observatory in Chile. PFS is a slit-fed echelle spectrograph with a wavelength coverage of $3910$--$7340$\,\AA. We used a 0.3\arcsec\ slit and $1 \times 2$\ binning, which yields a resolving power of $R \approx 133\,000$. Wavelength calibration is achieved via an iodine gas cell, which also allows the characterization of the instrumental profile. We obtained 20 spectra, observed through iodine, between 2019 Sep 12 and Oct 12, leading to six individual observation nights with typical exposure times of 20 minutes for each exposure. The last observations covered nine spectra and were meant to encompass the transit of TOI-1130\,b to detect the Rossiter--McLaughlin (RM) effect, but the observations missed the transit window due to TTVs. We also obtained an iodine-free template observation with an 80-minute exposure time. The radial velocities were extracted using a custom IDL pipeline following the prescriptions of \citetads{1992PASP..104..270M} and \citetads{1996ApJ...464L.153B}, and achieved a mean precision of 1.4\,\msunit. The velocities are presented in Table~\ref{tab:toi-1130_pfs}. Following the prescription of \citetads{2017AJ....153..208B}, we also calculated the S-index, measured from the core emission of the Ca {\sc ii} H \& K lines, and the H-index, which measures the chromospheric emission component in the H$\alpha$ line. These values are also reported in Table~\ref{tab:toi-1130_pfs}. 

\section{Data analysis}

The analysis is done in steps. First, we carry out stellar modeling of the spectra (Sect.~\ref{sec:iSpec} and \ref{sec:vosa}). We take advantage of the high-resolution and high signal-to-noise ratio (S/N) of the co-added HARPS spectra to independently derive the fundamental stellar parameters. Second, we perform a periodogram analysis of the radial velocities (Sect.~\ref{sec:frequency_analysis}) to identify which signals are present in the RV time series. Third, we modeled the photometry (Sect.~\ref{sec:transit_model}) to uncover the TTVs and extract the transit times \rev{used in the photodynamical analysis}. Each previous step is required to carry out the photodynamical modeling of the photometry and RV measurements (Sect.~\ref{sec:photo_model}) to determine the planet and orbital parameters.

\subsection{Stellar modeling with \ispec} 
\label{sec:iSpec} 

\begin{table}[!t]
\begin{threeparttable}[b]
 \centering
 \caption{Stellar parameters of TOI-1130.
 }
 \label{table:stellar_par}
	\begin{tabularx}{\linewidth}{@{\extracolsep{\fill}}lrr} 
    \toprule
    \toprule
Parameter &  & Reference \\
    \midrule
RA [$^\circ$] & 286.376006 & 1 \\
Dec [$^\circ$] & -41.43764 & 1 \\
Spectral type & K6--K7  & 2 \\
B [mag] & 12.42\,$\pm$\,0.26 & 3 \\
V [mag] & 11.59\,$\pm$\,0.16 & 3 \\
\tess [mag] & 10.1429\,$\pm$\,0.0061 & 4 \\
\gaia [mag] & 10.8989\,$\pm$\,0.0028 & 1 \\
J [mag] & 9.055\,$\pm$\,0.023 & 5 \\
H [mag] & 8.493\,$\pm$\,0.059 & 5 \\
K [mag] & 8.351\,$\pm$\,0.033 & 5 \\
WISE 3.4 um	[mag] & 8.266\,$\pm$\,0.022	& 6 \\		
WISE 4.6 um	[mag] & 8.339\,$\pm$\,0.019	& 6 \\		
WISE 12 um [mag] & 8.244\,$\pm$\,0.024 & 6 \\		
WISE 22 um [mag] & 8.472\,$\pm$\,0.361 & 6 \\
RUWE & 1.14 & 1 \\
Distance [pc] & $58.41\pm0.07$ & 1 \\
Age [Gyr] &  \rev{3.2--5} & this work\\
    \midrule
Parameter & \ispec \& \param & \vosa \\
    \midrule
\teff [K] & $4350 \pm 60$ & 4300--4400\\
\fehunit [dex] & $0.30 \pm 0.06$ & 0.0--0.5\\
\logg [cm\,$\mathrm{s}^{-2}$] & $4.62 \pm 0.04$ & 4--5\\
\vsini [\kms] & $\leq$\,3 & - \\
\smass [\msununit] & $0.71 \pm 0.02$ & - \\
\sradius [\rsununit] & $0.68 \pm 0.02$ & 0.66--0.74\tablefootmark{a}\\
\lum [$L_{\odot}$] & - & 0.148--0.154 \\
\bottomrule
\bottomrule
    \end{tabularx}
  \end{threeparttable}
  \tablefoot{
\tablefoottext{a}{Derived via Stefan–Boltzmann law.}
}
  \tablebib{
(1) \gaia eDR3 (\citeads{2016A&A...595A...1G}; \citeads{2022arXiv220800211G}; \citeads{2022arXiv220605989B}); (2) Empirical spectral type-colour sequence \citepads{2013ApJS..208....9P}; (3) Tycho-2 catalogue \citepads{2000A&A...355L..27H}; (4) TIC v8.2 \citepads{2022yCat.4039....0P}; (5) 2MASS \citepads{2003yCat.2246....0C}; (6) WISE \citepads{2014yCat.2328....0C}.
}
\end{table}

We used the co-added high-resolution HARPS spectrum (S/N\,$\approx$\,300 per pixel at 5500\,\AA) to determine the spectroscopic parameters of TOI-1130 using the \ispec framework (\citeads{2014A&A...569A.111B}; \citeads{2019MNRAS.486.2075B}). Specifically, we used the Spectroscopy Made Easy radiative transfer code (\sme; \citeads{1996A&AS..118..595V}; \citeads{2017A&A...597A..16P}), the MARCS atmospheres models \citepads{2008A&A...486..951G}, and the version 5 of the GES atomic line list \citepads{2015PhyS...90e4010H}, embedded in the \ispec framework. The models allow for fitting the effective temperature \teff between 2500--8000\,K, surface gravity \logg between 0.00--5.00\,dex, and metallicity \feh between $-$5.00--1.00\,dex. \ispec uses a nonlinear least-squares (Levenberg-Marquardt) fitting algorithm \citepads{2009ASPC..411..251M} to minimize the $\chi^2$ value between the observed spectra and the computed synthetic ones based on these models. We fitted simultaneously for an effective temperature, surface gravity, metallicity, and the projected stellar equatorial velocity in the region between 480 and 680\,nm. We used the empirical relations for the microturbulence and macroturbulence velocities ($V_{\rm mic}$, $V_{\rm mac}$) included in the \ispec framework to reduce the number of free parameters in our analysis. The spectral resolution was taken from spectrograph specifications. The effective temperature and metallicity derived in the \ispec analysis together with the \gaia eDR3 parallax and 2MASS J, H, K magnitudes were then put into the Bayesian parameter estimation code \param\footnote{\url{http://stev.oapd.inaf.it/cgi-bin/param}.} (\citeads{2006A&A...458..609D}; \citeads{2014MNRAS.445.2758R}, \citeyearads{2017MNRAS.467.1433R}). \param uses the PARSEC isochrones \citepads{2012MNRAS.427..127B} to estimate stellar parameters, such as stellar mass, radius, age, and surface gravity. We found good agreement between surface gravity derived in \ispec, but with reduced uncertainty. \param computes relative uncertainties assuming that the theoretical models represent reliable descriptions of stars. The whole procedure is described in \citetads{2006A&A...458..609D}.

\subsection{SED analysis with \vosa} 
\label{sec:vosa} 

As a \rev{reality} check, we also analyzed the spectral energy distribution (SED) with the Virtual Observatory SED Analyser (\vosa; \citeads{2008A&A...492..277B})\footnote{\url{http://svo2.cab.inta-csic.es/theory/vosa/}.}. We compared the results from five different models: BT-NextGen GNS93 (\citeads{1993A&A...271..587G}; \citeads{2006MNRAS.368.1087B}; \citeads{2012RSPTA.370.2765A}), BT-NextGen AGSS2009 (\citeads{2006MNRAS.368.1087B}; \citeads{2009ARA&A..47..481A}; \citeads{2012RSPTA.370.2765A}), BT-Settl-AGSS2009 (\citeads{2006MNRAS.368.1087B}; \citeads{2009ARA&A..47..481A}; \citeads{2012RSPTA.370.2765A}), BT-Settl-CIFIST (\citeads{2006MNRAS.368.1087B}; \citeads{2011SoPh..268..255C}; \citeads{2012RSPTA.370.2765A}), and Coelho Synthetic stellar library \citepads{2014MNRAS.440.1027C} to estimate \teff, \logg, \feh, and the interstellar extinction \Av. We use coordinates and distances from \gaia eDR3 and set limits in our analysis to \teff=\,3000--6000\,K, and \logg=\,4.0--5.0\,dex based on the parameters derived in Section \ref{sec:iSpec}. We set the limit for \feh between $-$0.5 and 0.5 dex if possible; however, the BT-NextGen GNS93 enables metallicity only up to 0.3, the Coelho Synthetic stellar library up to 0.2, and the BT-Settl-CIFIST are available only for the solar metallicity. For our SED fitting, we use the Tycho \citepads{2000A&A...355L..27H}, \gaia DR2 \citepads{2018A&A...616A...1G}, \gaia eDR3 \citepads{2021A&A...649A...1G}, 2MASS \citepads{2003yCat.2246....0C}, and WISE \citepads{2014yCat.2328....0C} photometry.

\vosa performs the ${\chi}^2$ minimization procedure to compare theoretical models with the observed photometry. For each model, we take three results with the lowest ${\chi}^2$. The lowest and highest value of each parameter from these results is used to create the final intervals. Additionally, the stellar radius is derived via the Stefan–Boltzmann law. 

We report the derived stellar parameters from the co-added HARPS spectra in Table \ref{table:stellar_par}.

\subsection{Periodogram analysis of the HARPS data}
\label{sec:frequency_analysis}

We searched the radial velocity data for the Doppler reflex motions induced by the two transiting planets. To this end, we performed a periodogram analysis of the HARPS and \rev{PFS} data \rev{as well as the} activity indicators (Sect.~\ref{HARPS_PFS_spectroscopy}), which also allowed us to search for the stellar rotational period and for possible additional Doppler signals that might be induced by the presence of other orbiting objects. 

\begin{figure}[!t]%
    \centering
    \includegraphics[width=\columnwidth]{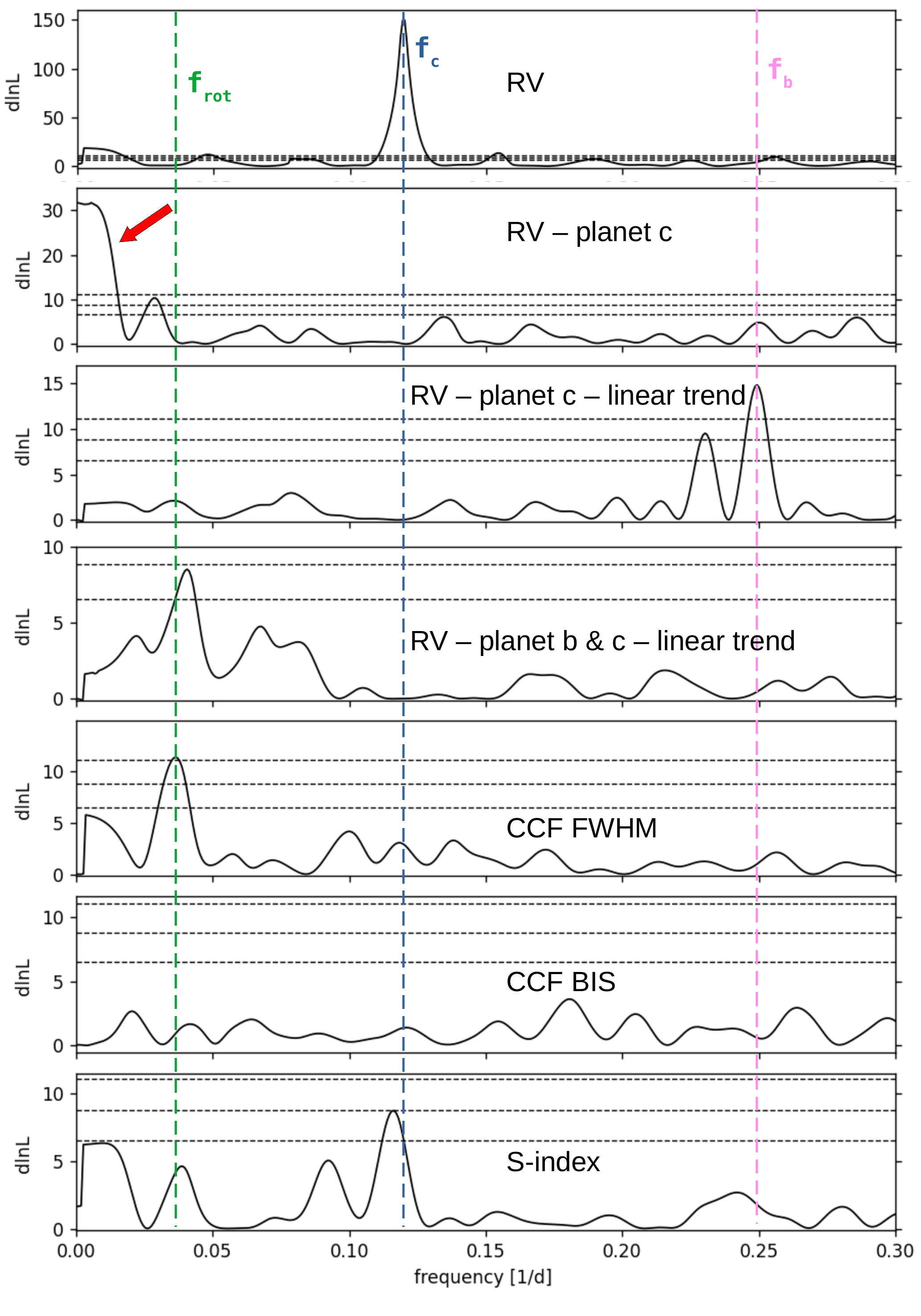} %
    \caption{\rev{Maximum likelihood periodograms of the HARPS+PSF RV measurements and HARPS activity indicators. The blue and salmon vertical dashed lines mark the orbital frequencies of the two transiting planets, and the green vertical dashed line marks the rotation period of the star, while the horizontal dashed lines mark the 10\,\%, 1\,\% and 0.1\,\% false alarm probability. From top to bottom: RV data; RV residuals after subtracting the signal of TOI-1130\,c; RV residuals after subtracting the signal of TOI-1130\,c and a linear trend; RV residuals after subtracting the signals of TOI-1130\,b \& c and a linear trend; FWHM and BIS of the CCF; S-index. The red arrow in the third panel marks the excess of power at low frequencies significantly detected in the HARPS RVs.}}
    \label{fig:gls_periodogram}%
\end{figure}
\rev{The results of the maximum likelihood periodogram (MPL) as implemented in \exostriker \citepads{2019ascl.soft06004T} are shown in Fig.~\ref{fig:gls_periodogram}. The HARPS and PFS RVs display a significant peak at the transit frequency of the outer giant planet TOI-1130\,c ($f_\mathrm{c}$\,$\approx$\,0.120\,days$^{-1}$; Fig.\,\ref{fig:gls_periodogram}, upper panel).}
 We subtracted the Doppler signal induced by TOI-1130\,c by fitting the \rev{RV} time series, assuming that the planet has a circular \rev{Keplerian} orbit. 
The \rev{MPL} periodogram of the RV residuals (second panel in Fig.~\ref{fig:gls_periodogram}) shows a significant excess of power at frequencies lower than \rev{$\sim$\,0.014\,days$^{-1}$}, the spectral resolution\footnote{Defined as the inverse of the baseline, i.e., \rev{1/70\,=\,0.014\,days$^{-1}$, where 70 days is the baseline of our RV follow-up.}} of our \rev{RV} data, indicating the presence of a long-term trend in our Doppler measurements (red arrow in Fig.~\ref{fig:gls_periodogram}, \rev{second} panel). This peak has no counterpart in any of the periodograms of the activity indicators, suggesting that it \rev{could potentially be produced by an} additional outer orbiting companion (see Sect.~\ref{sec:results}). 
\rev{After subtracting the previously detected signals, the residuals shown in the third panel of Fig.~\ref{fig:gls_periodogram} show a significant peak at the transit frequency of the inner planet TOI-1130\,b ($f_\mathrm{b}$\,$\approx$\,0.245\,days$^{-1}$)}. This peak is not significantly detected in any of the activity indicators, namely the CCF FWHM and BIS, and the S-index (Fig.~\ref{fig:gls_periodogram}, fifth, sixth, and seventh panel), spectroscopically confirming that the transit signal detected by \tess\ is due to planet\,b. The RV residuals following the subtraction of the Doppler motion induced by the two transiting planets and the linear trend (Fig.~\ref{fig:gls_periodogram}, fourth panel), display a peak at $\sim$\,0.041\,days$^{-1}$ (i.e., $\sim$\,24.4\,days) with a false alarm probability of \rev{FAP\,$\approx$\,1\,\%}. Although the RV residuals' peak is insignificant, we note that \rev{this peak and the peak in the CCF FWHM ($\sim$\,0.0.037\,days$^{-1}$; Fig.~1, fifth panel) with a FAP of $\approx$\,0.1\,\%} are virtually indistinguishable. Their difference of 0.004\,days$^{-1}$ is \rev{3.5} times smaller than the spectral resolution of our RV time-series \rev{(0.014\,days$^{-1}$)}. Given the \rev{K6-K7} spectral type of TOI-1130 (Table~\ref{table:stellar_par}), the signal at $\sim$\,27 days is very likely due to the presence of photospheric active regions carried around by stellar rotation. As such, we interpret the signal at $P_\mathrm{rot}$\,$\approx$\,27\,days \rev{as the star's rotation period}. The S-index shows a peak close to the orbital frequency of planet c, but with a FAP of \rev{1\,\%} it is not significant.



\subsection{Photometric modeling}
\label{sec:transit_model}

\begin{figure*}[!t]%
    \centering
    \includegraphics[width=0.99\textwidth]{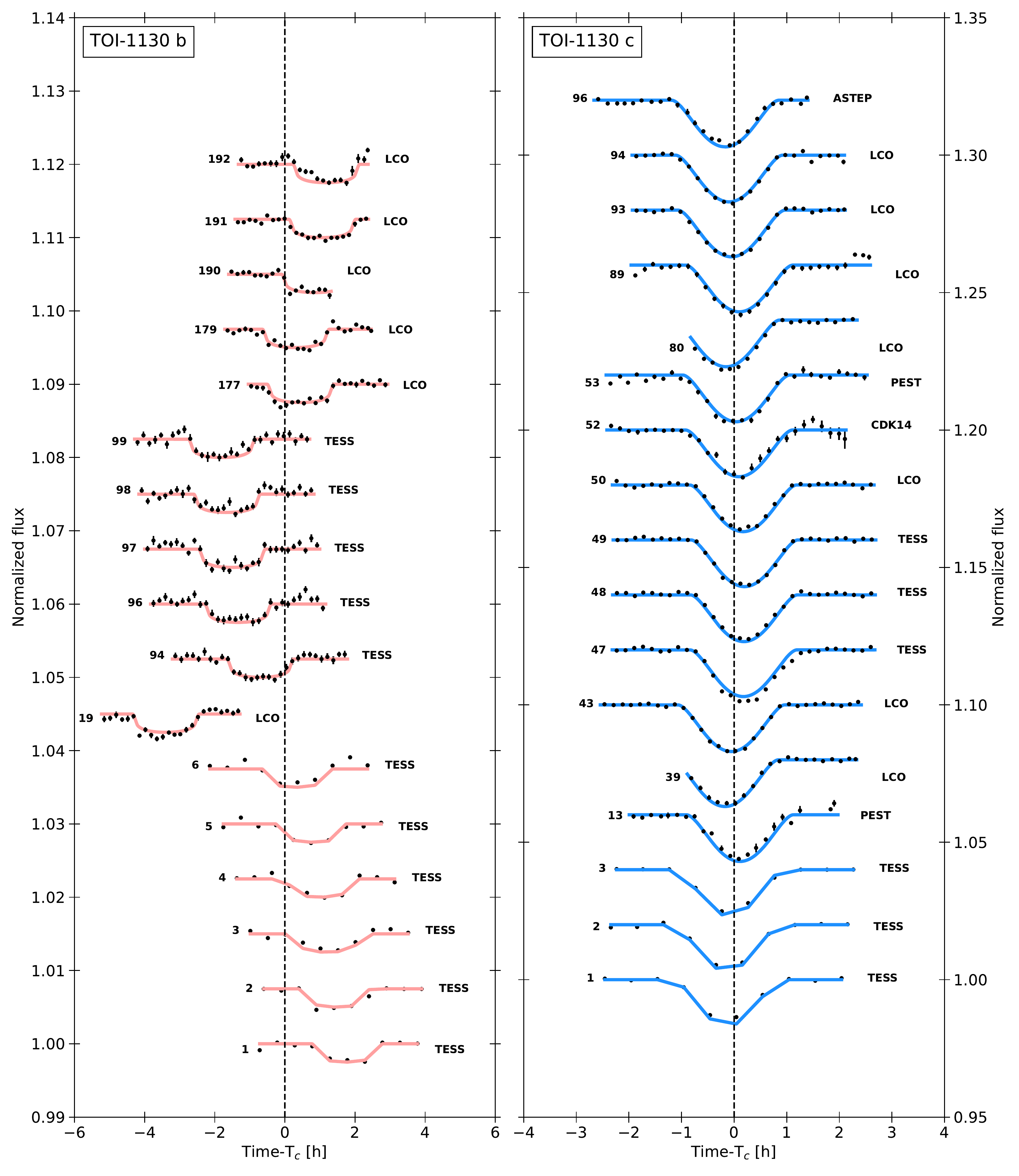} %
    \caption{Detrended \tess and ground-based light curves for TOI-1130\,b (left) and for TOI-1130\,c (right). The light curves are shifted in phase according to the derived linear mean orbital period of each planet and shifted arbitrarily vertically for visibility. The corresponding transit number is shown on the left of each transit. For more details see Table~\ref{table:ground-obs-table}. All light curves are binned to 10-min, except for the 30-min \tess light curves. The best-fit \pyttv transit model (see Sect.~\ref{sec:transit_model}) is overplotted (TOI-1130\,b: salmon and TOI-1130\,c: blue). Note the different y-scales.}%
    \label{fig:ttv_plot}%
\end{figure*}

We fit the previously observed \tess and ground-based transits reported in \citetads{2020ApJ...892L...7H} together with the new observations from \tess and the additional ground-based observations reported here using the Python Tool for Transit Variation \citep[\pyttv;][]{kups11289}. \rev{This fit was done to first test whether the system shows TTVs and then to extract the transit center times for the photodynamical analysis. Other quantities were of no interest in this step.}

\pyttv can search for and identify transit variations in light curves, and fit transits accounting for possible variations in the orbital period. By this method, the transits from both planets are modeled simultaneously using the quadratic \citetads{2002ApJ...580L.171M} transit model with the Taylor-series expansion from \citetads{2020MNRAS.499.3356P} as implemented in \pytransit \citepads{2015MNRAS.450.3233P}. For the 30-min cadence data from Sector~13, the transit model is super-sampled as suggested by \citetads{2010MNRAS.408.1758K} to ensure a robust transit fit. The fit uses the orbital period \p, the epoch \epoch, the planet radius relative to stellar radius \rpoverrstar~$\equiv$\,\kp, the transit center times \tc, and impact parameter \pimp for each planet as free parameters. Shared parameters during the fit are the quadratic limb darkening parameters \qone, \qtwo as introduced in \citetads{2013MNRAS.435.2152K}, and the stellar density \rhostar. We give each pass-band the same limb-darkening coefficients because limb darkening has little effect on the center time estimates. To account for stellar activity, we modeled the baseline as a Gaussian Process (GP) with a Mat\'ern 3/2 kernel as implemented in \celerite \citepads{2017AJ....154..220F}. 
We set wide normal priors on the orbital periods, the epochs, and the transit depths, where the prior means correspond to the values reported in ExoFOP-TESS\footnote{\url{https://exofop.ipac.caltech.edu/tess/target.php?id=254113311}}. \rev{These priors are only weakly informative and do not constrain the posteriors in any significant way. They only aid the global optimizer.}
For the other parameters, we used uniform priors. We estimated the parameter posteriors using Markov Chain Monte Carlo (MCMC) sampling as implemented in \emcee \citepads{2013PASP..125..306F}. 

\subsection{Photodynamical modeling}
\label{sec:photo_model}

\begin{table*}[!t]
\begin{threeparttable}
    \centering
    \small
    \caption{Parameters of the TOI-1130 system. The values and the 1$\sigma$ uncertainties for planets in the TOI-1130 system are estimated through the photodynamical modeling using \pyttv. The reported osculating orbital elements are valid for the reference time \epoch\,=\,2458657.}
    \label{tab:toi-1130_values}
    \begin{tabularx}{\textwidth}{@{\extracolsep{\fill}}lrrrr}
    \toprule
    \toprule
     & \multicolumn{4}{c}{TOI-1130}\\
    \\
    \textit{Fitted stellar parameters} & & Prior & Posterior & \\
    \midrule
    \sradius [\rsun] & & \NP{0.68,0.02} & 0.695\,$\pm$\,0.015 & \\
    \smass [\msun] & & \NP{0.71,0.02} & 0.712\,$\pm$\,0.017 & \\
    ${q}_1$ & & \UP{0,1} & 0.65\,$\pm$\,0.27 & \\
    ${q}_2$ & & \UP{0,1} & 0.31\,$\pm$\,0.21 & \\
    RV slope [m\,$\mathrm{s}^{-1}$\,$\mathrm{days}^{-1}$] & & \NP{0,1} & 0.495\,$\pm$\,0.021 & \\
    $\gamma_\mathrm{HARPS}$ [m\,$\mathrm{s}^{-1}$] & & \NP{-7959,90} & -7968.35\,$\pm$\,0.37 & \\
    $\gamma_\mathrm{PFS}$ [m\,$\mathrm{s}^{-1}$] & & \NP{32,90} & 93.79\,$\pm$\,0.78 & \\
    RV-jitter$_\mathrm{HARPS}$[m\,$\mathrm{s}^{-1}$] & & \NP{0,0.5} & 0.15\,$\pm$\,0.16 & \\
    RV-jitter$_\mathrm{PFS}$ [m\,$\mathrm{s}^{-1}$] & & \NP{0,0.5} & 0.37\,$\pm$\,0.10 & \\
    $P_\mathrm{rot}$ [days] & & \NP{27,2} & 25.6\,$\pm$\,1.2 & \\
    \\
    \textit{Derived stellar parameters} & & & & \\
    \midrule
     $\rho_\star$ [g\,$\mathrm{cm}^{-3}$] &  & - & 2.98\,$\pm$\,0.18 & \\
    \\
     & \multicolumn{2}{c}{TOI-1130\,b} &\multicolumn{2}{c}{TOI-1130\,c}\\
     \\
    \textit{Fitted planet parameters} & Prior & Posterior & Prior & Posterior \\
    \midrule
    \p [days] & \NP{4.07,0.02} & 	
    4.07445\,$\pm$\,0.00046 & \NP{8.35,0.02} & 8.350231\,$\pm$\,0.000098\\
    \epoch [BJD] & \NP{2458658.738, 0.003} & 2458658.7405\,$\pm$\,0.0013 & \NP{2458657.798, 0.003} & 2458657.90322\,$\pm$\,0.00030\\
    $\log_{10}$\,$M_\mathrm{p}$ [$\log_{10}\msun$] & \UP{-5.5, -3.5} & -4.237\,$\pm$\,0.022 &  	
    \UP{-4, -2} & -3.0096\,$\pm$\,0.0074\\ 			
    \rpoverrstar & \UP{0.03, 0.06} & 0.0470\,$\pm$\,0.0011 & \UP{0.09, 0.25} & 0.175$^{+0.018}_{-0.016}$\\ 	
    $\sqrt{e}\cos{\omega}$ & \UP{-0.5, 0.5} & 0.1358\,$\pm$\,0.0021 & \UP{-0.5, 0.5} & -0.010\,$\pm$\,0.022\\ 
    $\sqrt{e}\sin{\omega}$ & \UP{-0.5, 0.5} & -0.1889\,$\pm$\,0.0037 & \UP{-0.5, 0.5} & -0.2124\,$\pm$\,0.0040\\ 	
    \pimp & \UP{-1, 1} & -0.518\,$\pm$\,0.054 & \UP{-1, 1} & -0.941$^{+0.030}_{-0.032}$\\
    \Omegaorbit [rad] & $\pi$ & $\pi$ & \UP{0.5\pi,1.5\pi} & 3.1232\,$\pm$\,0.0066\\
    \\
    \textit{Derived parameters} & & & \\
    \midrule
    $M_\mathrm{p}$ [\mearth] &  & 19.28\,$\pm$\,0.97 &  & 325.69\,$\pm$\,5.59 \\ 	
    $R_\mathrm{p}$ [\rearth] &  & 3.56\,$\pm$\,0.13 & & 13.32$^{+1.55}_{-1.41}$\\	
    $\rho_\mathrm{p}$ [g\,$\mathrm{cm}^{-3}$] &  & 2.34\,$\pm$\,0.26 & &  0.75$^{+0.31}_{-0.21}$\\
    $e$ & \NP{0, 0.083} & 0.0541\,$\pm$\,0.0015 & \NP{0, 0.083} &  0.0457\,$\pm$\,0.0016\\
    $\omega$ [$^\circ$] & & 144.24\,$\pm$\,0.72 & & 175.23\,$\pm$\,4.24\\			
    $i$ [$^\circ$]\tablefootmark{a} & & 92.14\,$\pm$\,0.26 & & 92.44\,$\pm$\,0.14\\ 	
    $a/R_\star$ & & 13.77\,$\pm$\,0.27 & & 22.22\,$\pm$\,0.44\\
    $a$ [AU] & & 0.04457\,$\pm$\,0.00036 & & 0.07191\,$\pm$\,0.00058\\
   $T_\mathrm{eq}$\tablefootmark{b} [K] & & 632.17\,$\pm$\,12.60 & & 497.70\,$\pm$\,9.92 \\
   $F_\mathrm{p}$ [$F_\oplus$] & & 78.10\,$\pm$\,5.55 & & 30.00\,$\pm$\,2.13 \\ 
        \bottomrule
        \bottomrule
            \end{tabularx}
           \end{threeparttable}
           \tablefoot{
\tablefoottext{a}{Note the degeneracy in orbit inclination.}
\tablefoottext{b}{Calculated following equation (2) from \citetads{2005ApJ...626..523C} with f = 1 and bond albedo of 0.3.}
}
\end{table*}
Since the planets strongly interact gravitationally with each other to produce significant TTVs as shown by the photometric modeling in the previous section (Fig.~\ref{fig:ttv_plot}), we decided to model the RVs and the photometry simultaneously using a photodynamical model to determine the final planetary and orbital parameters. A photodynamical model combines a photometric transit model with N-body simulations whereby it produces a light curve that can be compared with observed light curves. The advantage of a photodynamical model is that it models the transit light curves taking into account the gravitational interaction between the bodies in the system. It models the transits simultaneously and includes also the transit shapes rather than only individual center times compared to traditional TTV analysis. This results in more robustly determined parameters as pointed out by several authors (e.g., \citeads{2018MNRAS.478..460A}). The disadvantage, however, is the high computational cost a photodynamical model needs. Here, we used the photodynamical model as implemented in \pyttv assuming a two-planet model, a sinusoidal RV signal to account for the stellar rotation, and a linear RV trend following the results from the periodogram analysis in Sect.~\ref{sec:frequency_analysis}. We briefly summarize the main parts of the photodynamical model.

The model uses the transit model with quadratic limb-darkening law described in \citetads{2015MNRAS.450.3233P} with the fast orbit computation introduced in \citetads{2020MNRAS.499.3356P}, and the numerical N-body code \rebound with the IAS15 integrator (\citeads{2012A&A...537A.128R}; \citeads{2015MNRAS.446.1424R}) and the library \reboundx \citepads{2020MNRAS.491.2885T}, to fit all the parameters without approximations. The transit model is super-sampled with ten samples per exposure for the long cadence observations (30-min), while short cadence observations (2-min) are calculated with one sample. We decided to fit the 2-min cadence instead of the 20-sec cadence because the 20-sec cadence would not lead to better time precision due to the relatively low S/N of the individual transits. The photodynamical model is parameterized by the stellar mass \smass and radius \sradius, the logarithms of the planet masses, planet radii relative to stellar radius \rpoverrstar~$\equiv$\,\kp, the impact parameters \pimp, the quadratic limb darkening parameters \qone, \qtwo defined in \citetads{2013MNRAS.435.2152K}, and the orbital elements (\rev{transit center time \epoch}, orbital period \p, eccentricity \ecc, argument of periastron \omegaorbit where \ecc and \omegaorbit are mapped from sampling parameters $\sqrt{e}\cos{\omega}$ and $\sqrt{e}\sin\omega$, longitude of the ascending node \Omegaorbit) at a reference time. The RV part adds further parameters, a linear trend, one offset for each telescope, three parameters for the sinusoidal function modeling the stellar activity (amplitude, period, and phase), \rev{and an RV-jitter term for each RV-data set.}. The photometric variability is again modeled \rev{as} a GP with a Mat\'ern 3/2 kernel using \celerite. The estimation of physical quantities is carried out via a global optimization followed by MCMC sampling using \emcee to obtain samples from the posterior. 

We fitted the \tess photometry from both sectors, the transit center times for the ground-based transits estimated in Sect.~\ref{sec:transit_model}, and the RVs simultaneously with the photodynamical model. We decided to fit only for the ground-based transit center times rather than the full transits to simplify the process.\footnote{Fitting each ground-based transit with the photodynamical model complicates the analysis due to the different band-passes, noise properties, and exposure times.} We did not include the RV observed with CHIRON published in \citetads{2020ApJ...892L...7H} since the RV \rev{uncertainties are 10 times larger for CHIRON than for HARPS or PFS (20~m\,$\mathrm{s}^{-1}$ vs. 2~m\,$\mathrm{s}^{-1}$)}. The photodynamical model was initialized with the values reported in Table~\ref{tab:toi-1130_values} and has in total \rev{28} free parameters: stellar mass and radius, six orbital elements, one mass and one radius ratio for each planet, two limb darkening coefficients, an RV trend, one RV offset for each telescope, amplitude, period and phase of the sinusoidal function, \rev{and two RV-jitter terms}. We included an RV trend in the modeling to account for the Doppler signal induced by the \rev{potential} outer companion detected in the HARPS data (Sect.~\ref{sec:frequency_analysis}). We decided to model this third signal as a trend given the relatively short baseline of our RV follow-up. The periodogram analysis (Sect.~\ref{sec:frequency_analysis}) also revealed the stellar rotational period. We thus decided to account for the stellar rotation signal in the RVs by including a sinusoidal function in the RV model. The period of the signal was loosely constrained by a normal prior (see Table~\ref{tab:toi-1130_values}), while its phase and amplitude were given non-constraining uniform priors.

\rev{The photodynamical modeling allows us to constrain the mutual inclination between the planets which is calculated via:
\begin{equation}
    I = \cos^{-1}{(\cos{i_\mathrm{b}}\cos{i_\mathrm{c}}+\sin{i_\mathrm{b}}\sin{i_\mathrm{c}}\cos{(\Omega_\mathrm{b}-\Omega_\mathrm{c})})}.
\end{equation}}

Even if both planets were detected significantly in the RVs, we studied the effect of each observation method separately. Thus, we ran two additional photodynamical models where we used either only the RV measurements or the photometric observations to determine the system parameters. 

\section{Results}
\label{sec:results}

\begin{figure}[!t]%
    \centering
    \includegraphics[width=\columnwidth]{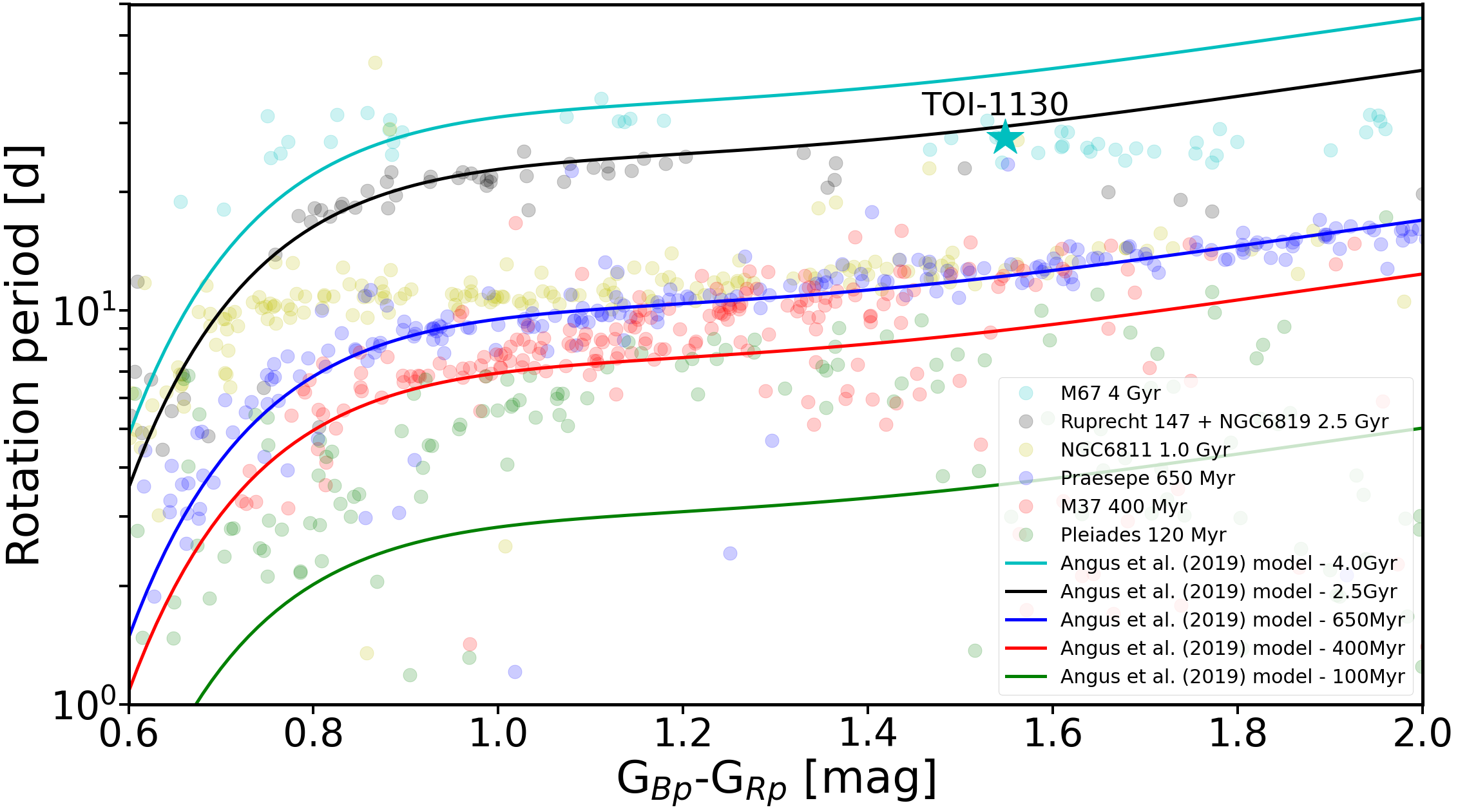} %
    \caption{Gaia $B_{P}-R_{P}$ colour vs. rotation period diagram for TOI-1130 (cyan star) and members
of well-studied clusters. Lines represent the empirical relation calibrated on the Praesepe cluster from \citetads{2019JOSS....4.1469A}.}
    \label{fig:age}%
\end{figure}

\begin{figure*}[t]%
    \centering
    \includegraphics[width=0.49\textwidth]{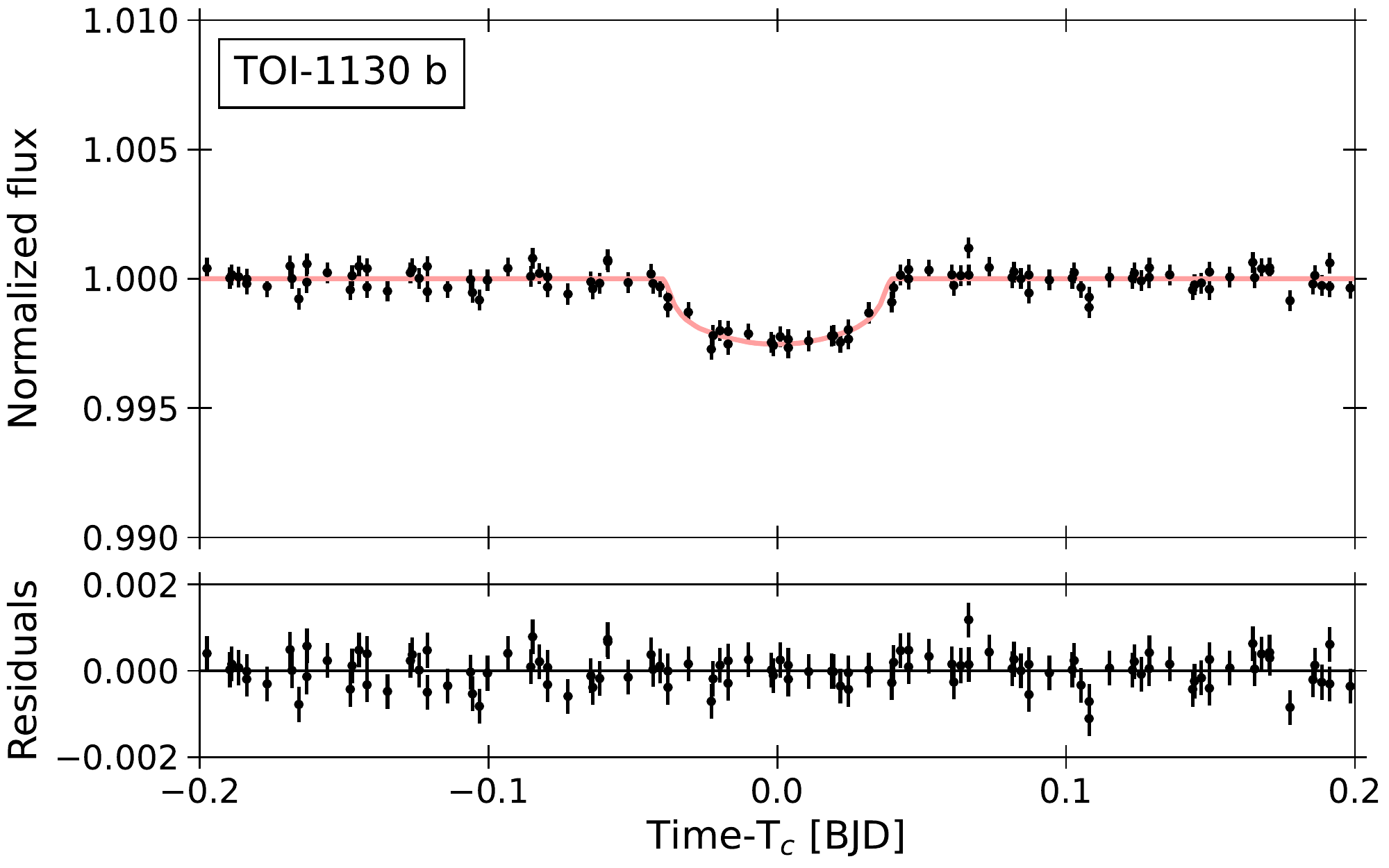} %
    \includegraphics[width=0.49\textwidth]{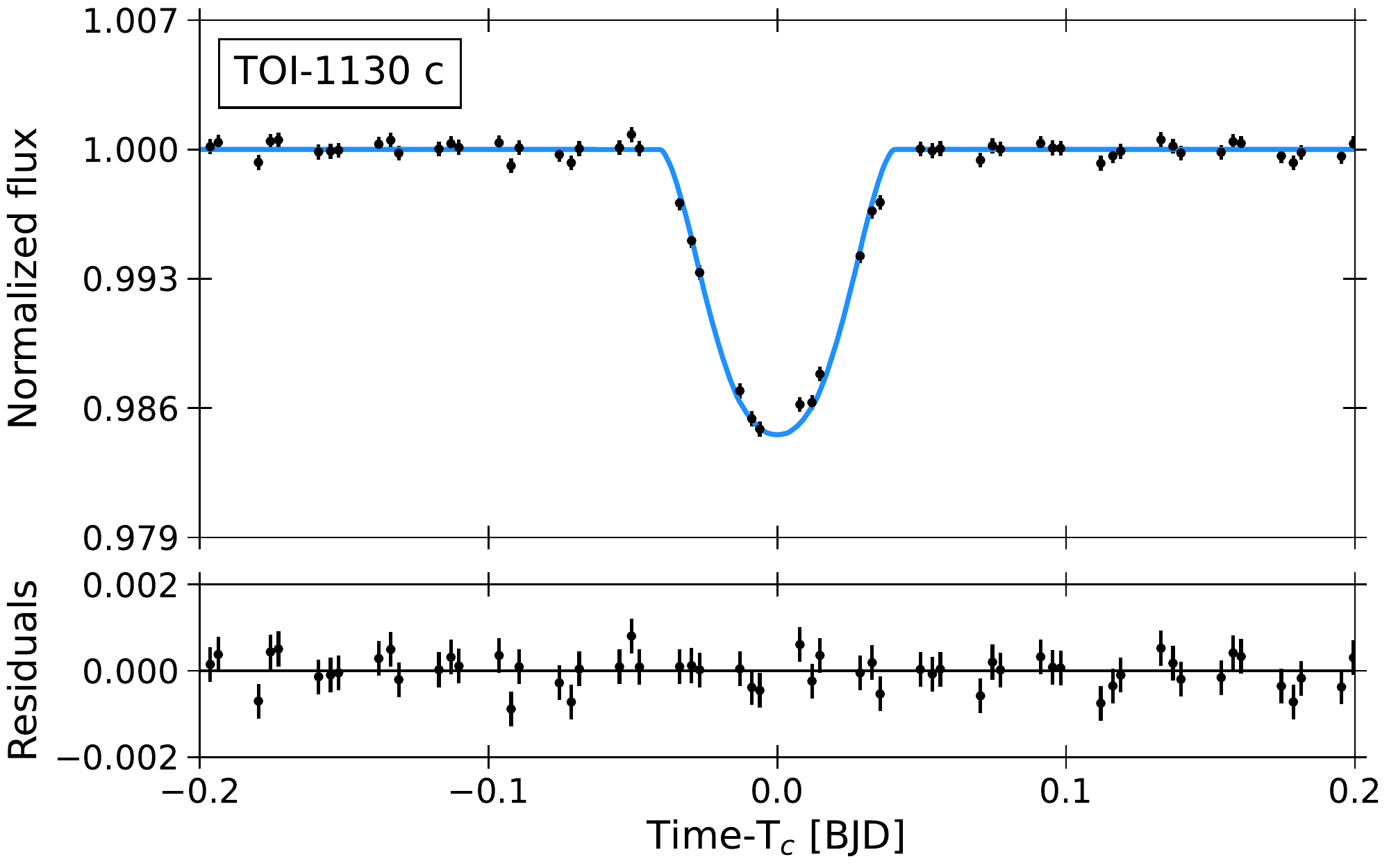}%

    \includegraphics[width=0.49\textwidth]{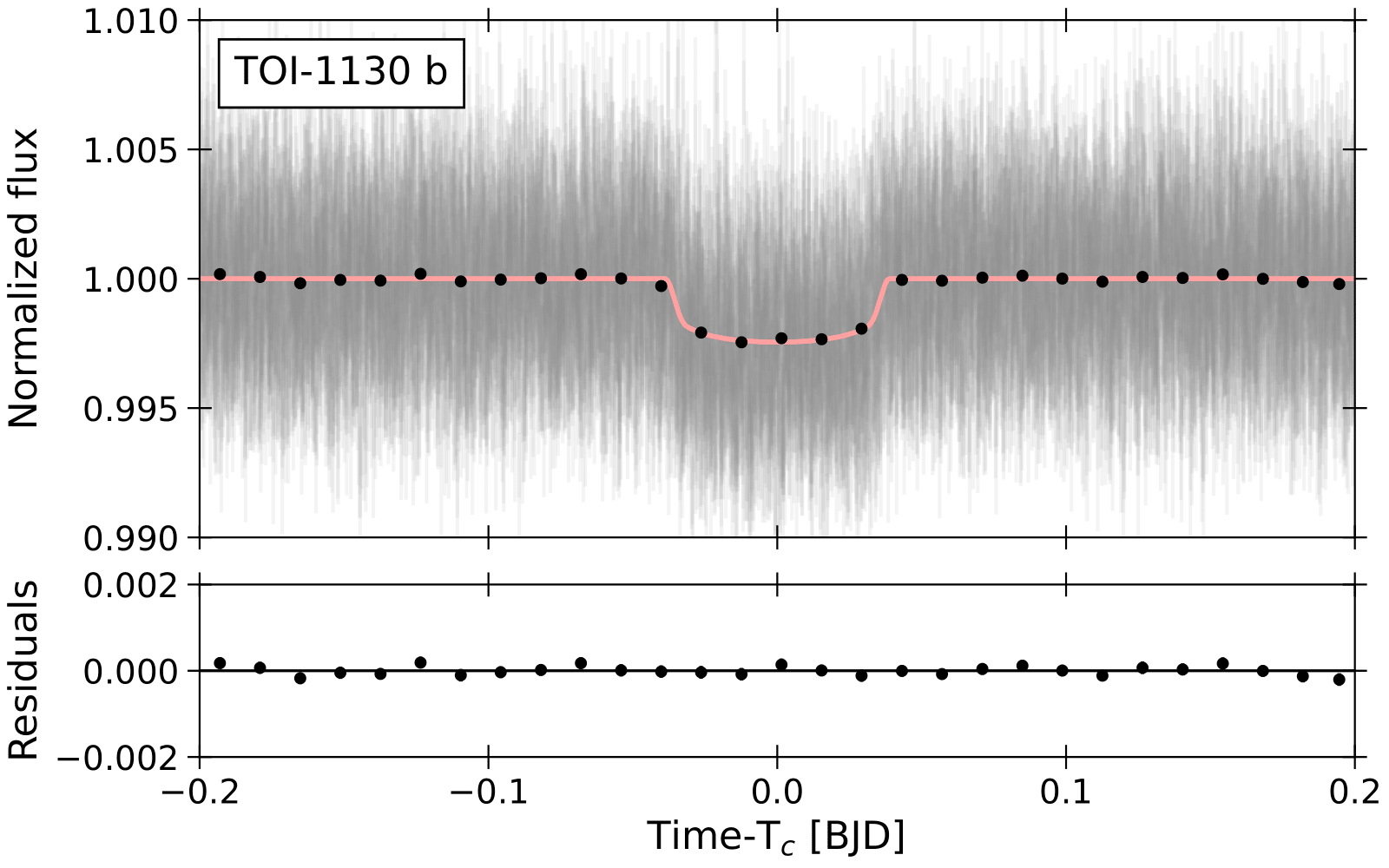} %
    \includegraphics[width=0.49\textwidth]{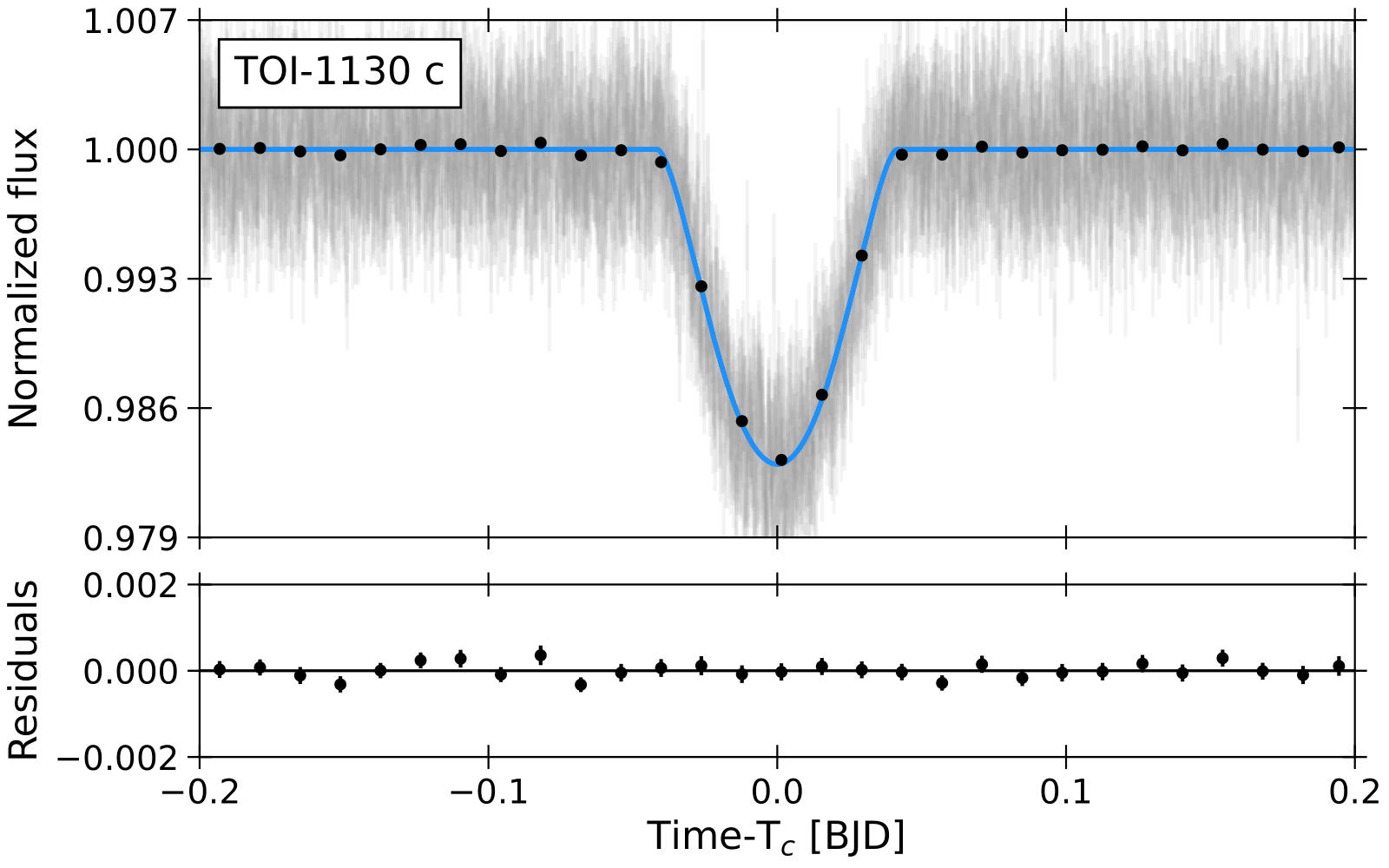}%
    \caption{Phase-folded \tess transit light curves for TOI-1130\,b (left) and for TOI\,1130\,c (right) accounting for the transit timing variations. The best-fit \pyttv transit model (see Sect.~\ref{sec:transit_model}) is overplotted for both TOI-1130\,b and TOI-1130\,c color-coded in salmon and blue, respectively. The upper panels show the light curves from Sector~13 observed in the 30-min cadence mode (black points) and the lower panels show the light curves from Sector~27 observed in the 20-s cadence mode (grey points) with a binning of 20-min (black points). Residuals are shown beneath each plot.}%
    \label{fig:tess_phase_folded}%
\end{figure*}

\begin{figure*}[t]%
    \centering
    \includegraphics[width=\textwidth]{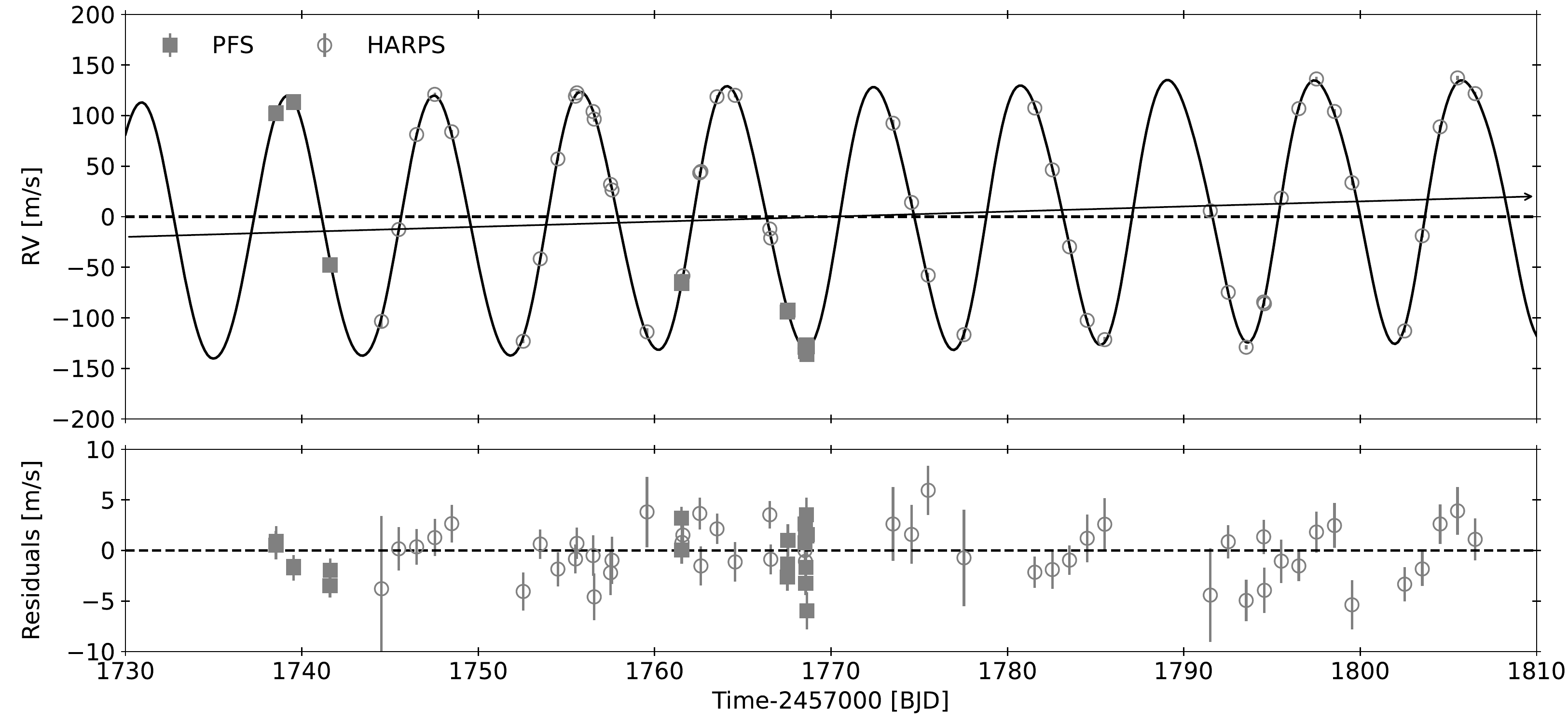} %
    \includegraphics[width=\textwidth]{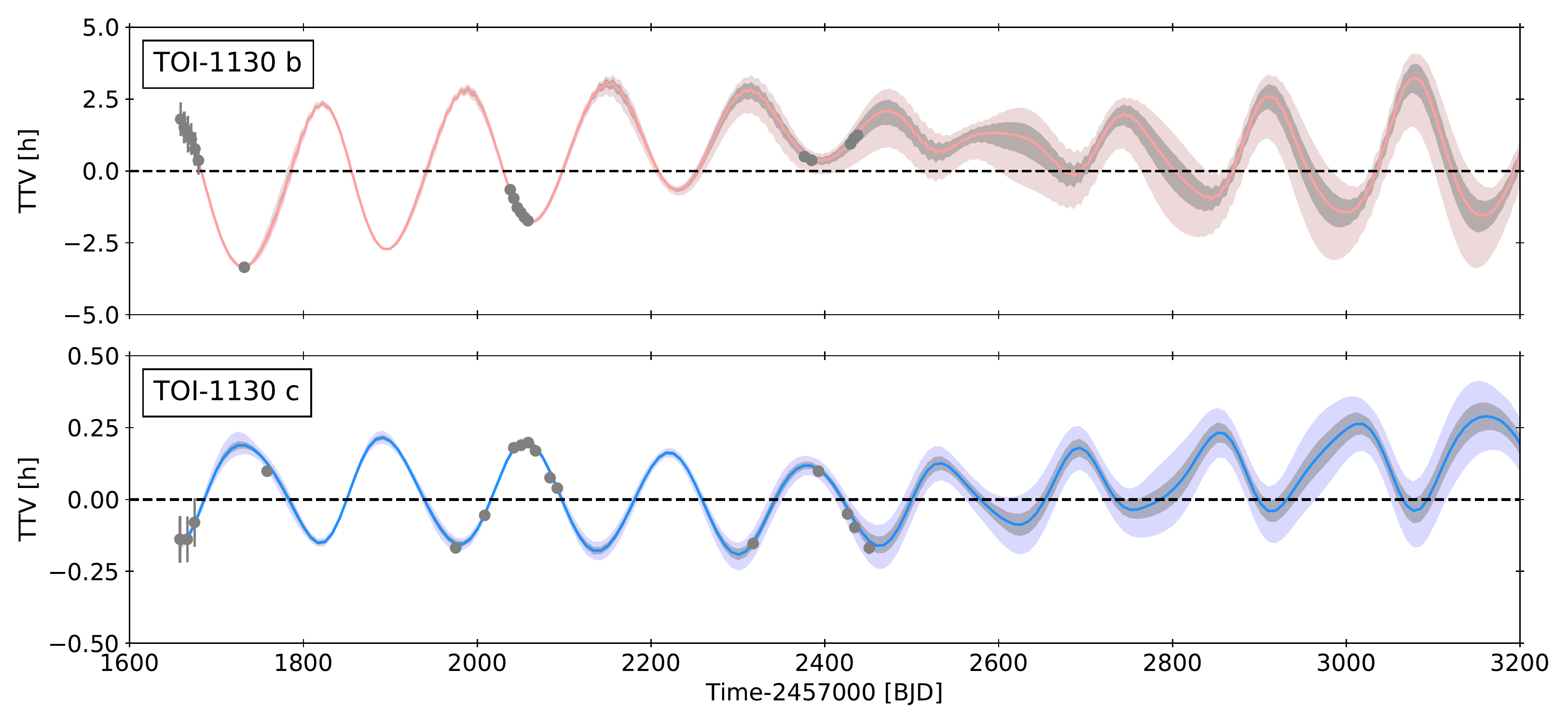} %
    \caption{RV time series (upper panel) for TOI-1130 where the open circles and filled squares mark the data from HARPS and PFS, respectively. The best-fit \pyttv model (see Sect.~\ref{sec:photo_model}) is overplotted in black. The data show a clear linear trend which is marked by the black arrow. RV residuals are shown in the second panel. Posterior TTV model (lower panels) from the photodynamical modeling of TOI-1130\,b (third panel) and TOI-1130\,c (fourth panel) with \pyttv (see Sect.~\ref{sec:photo_model}). The solid salmon and blue colored lines mark the median of the posterior pdf while the shaded areas in grey and in light blue and pink show the 1\,$\sigma$ and 3\,$\sigma$ ranges for TOI-1130\,b and TOI-1130\,c, respectively. The TTVs with their individual uncertainties measured by fitting each transit center as an independent free parameter with \pyttv (Sect.~\ref{sec:transit_model}) are shown in grey for comparison. The dashed lines mark the subtracted mean orbital period of 4.079$\pm$0.022\,days with an ephemeris of 2458658.66924\,$\pm$\,0.00020\,days, and of 8.3495$\pm$0.0027\,days with an ephemeris of 2458657.910329\,$\pm$\,0.000046\,days, respectively for TOI-1130\,b and c. 
    }%
    \label{fig:rv_plot}%
\end{figure*}

\begin{figure}[!h]%
    \centering
    \includegraphics[width=\columnwidth]{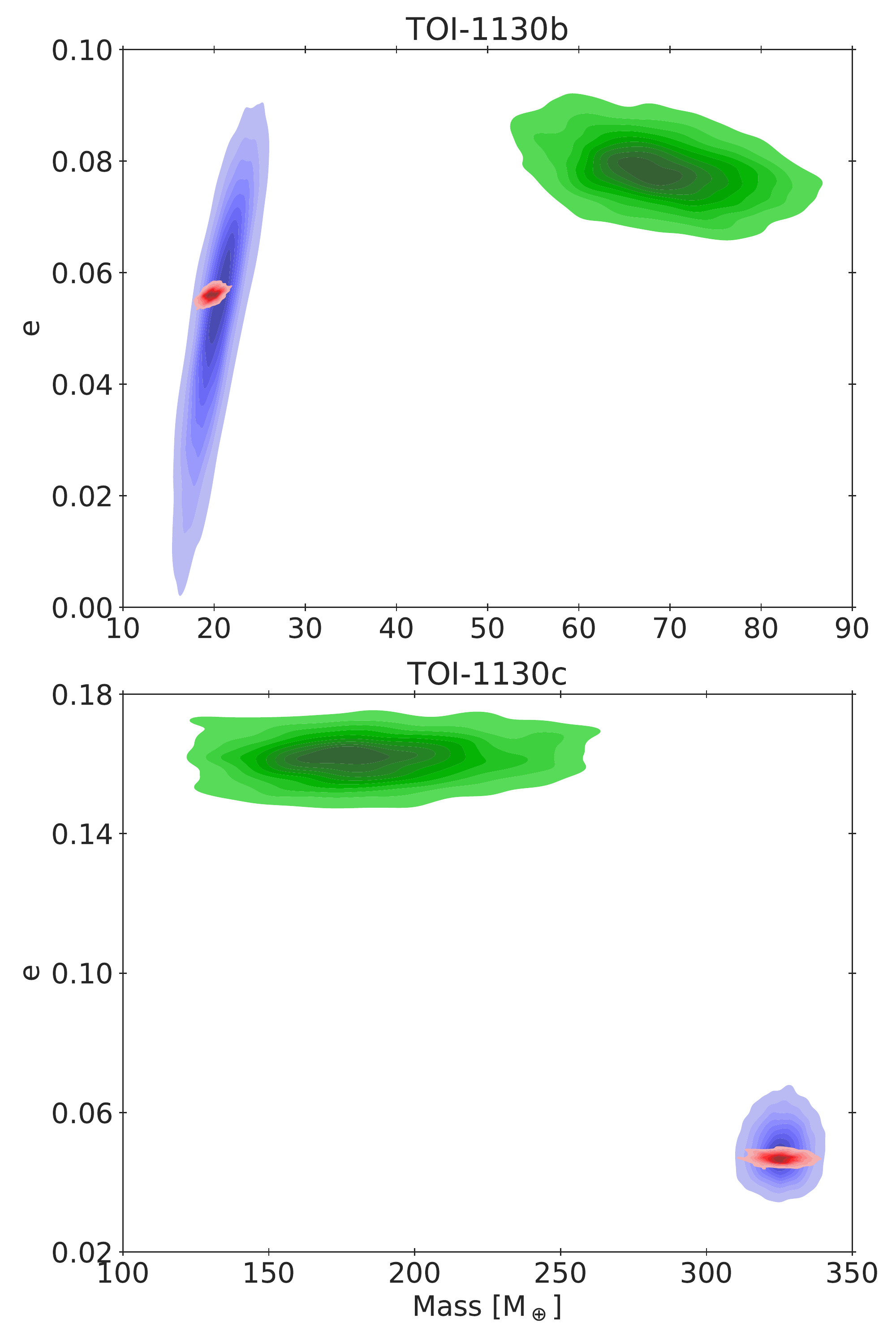} %
    \caption{Eccentricity-mass-contour plot of the posteriors from the photodynamical model fitting RVs (blue) and TTVs (green) separately and simultaneously (red).}%
    \label{fig:ttv_rv_contour}%
\end{figure}

The values derived from \ispec and \param agree with those inferred from \vosa and are adopted as the final stellar parameters. 

We found that the light of TOI-1130 suffers a negligible reddening, as expected given the proximity of TOI-1130 to the Sun ($d=58.41\pm0.07$\,pc, Table~\ref{table:stellar_par}). The V-band interstellar extinction \Av ranges between 0 and 0.05, depending on the atmospheric model used to fit the spectral energy distribution. As a \rev{reality} check we modeled the SED following the method described in \citetads{Gandolfi2008} and found an \Av~=\,0.03\,$\pm$\,0.03, in agreement with the \vosa results. 

The derived stellar parameters agree within 1$\sigma$ with the ones reported in \citetads{2020ApJ...892L...7H} derived from SED fitting. \citetads{2020ApJ...892L...7H} estimated a \vsini=\,4.0\,$\pm$\,0.5\,\kms\, from the light curve while our \vsini$<$\,3\,\kms\, estimated from \ispec is lower.

Our conservative stellar age determination of $4.86^{+3.38}_{-4.26}$\,Gyr from stellar isochrones agrees within 1$\sigma$ with the lower end of the age $8.2^{+3.8}_{-4.9}$\,Gyr derived by \citetads{2020ApJ...892L...7H}. Nevertheless, the age determination using isochrones of K dwarfs such as TOI-1130 is not reliable, since the star has already reached the main sequence. The periodogram analysis in Sect.~\ref{sec:frequency_analysis} indicates a stellar rotational period of $\approx$\,27\,days, which was confirmed in the photodynamical analysis to be \rev{25.6\,$\pm$\,1.2\,days}. 
To better constrain the system's age, we compared the rotation period of TOI-1130 with the gyrochronology empirical relation from \citetads{2019JOSS....4.1469A} calibrated on the Praesepe cluster and the members of well-studied clusters \rev{studied in \citeads{Godoy21}: Pleiades cluster ($\sim$120\,Myr), M37 cluster ($\sim$400\,Myr), Praesepe cluster ($\sim$650\,Myr), NGC\,6811 cluster ($\sim$1\,Gyr), together with the Ruprecht\,147\,+\,NGC\,6819 clusters ($\sim$2.5\,Gyr) studied in \citeads{Curtis20} and the M67 cluster ($\sim$4\,Gyr) studied in \citeads{Dungee22}. Each cluster was corrected for an interstellar extinction using the {\tt dustmaps} code \citepads{Green18} and three-dimensional Bayestar dust maps \citepads{Green19}.} 
\rev{We can see in Fig.~\ref{fig:age} that the system's age is consistent with the M67 cluster. Hence, we adopted the final age as an age of M67 from the literature. The age 3.2--5\,Gyr is in agreement with the age derived by isochrones fitting.}

We show the phase-folded transits from Sector~13 and 27 separately in Fig.~\ref{fig:tess_phase_folded} and the ground-based transits observed with the different facilities in Fig.~\ref{fig:ttv_plot} along with the phase-folded best-fit model from \pyttv (see Sect.~\ref{sec:transit_model}). As seen from Fig.~\ref{fig:ttv_plot}, both planets show significant TTVs as already suspected in \citetads{2020ApJ...892L...7H}. The newly observed \tess and ground-based transits confirm their expected TTV amplitude of at least two hours for planet b. 

The orbital and planetary parameters are derived from the photodynamical modeling that fits the RV, the \tess photometry, and the \rev{additional} ground-based transit times simultaneously (Sect.~\ref{sec:photo_model}) \rev{and reported in Table.~\ref{tab:toi-1130_values}}. The RV model derived from the photodynamical fit to the HARPS and PFS RVs, the \tess photometry, as well as the ground-based transit center times, is shown in the upper panel of Fig.~\ref{fig:rv_plot}, while the TTV model is shown in the lower panels of Fig.~\ref{fig:rv_plot}. We show the individual photodynamical transit models to the \tess photometry in Fig.~\ref{fig:transits_photodyn_b}. 

We show the eccentricity--mass contour plot of the photodynamical model fitting only the RVs, the TTVs, and the RVs and TTVs simultaneously in Fig.~\ref{fig:ttv_rv_contour}, to understand the contribution of both spectroscopic and photometric observations. \rev{The posterior using only the photometric data lies far from the posteriors fitting only the RVs and RV and TTV simultaneously. This difference is caused by insufficient coverage of the TTV phase. The planetary masses and orbital eccentricities cannot be derived from the current photometric data due to known mass-eccentricity degeneracy. The photometric solution fits a different posterior mode.} Fitting only the RVs constrains the planetary masses tightly but leaves the eccentricity for the inner planet unconstrained. The photodynamical model that fits both the RVs and TTVs simultaneously allows us instead to put tight constraints on both the orbital eccentricities and planet masses of both planets. 

The RV time series in the upper panel in Fig.~\ref{fig:rv_plot} shows a significant linear trend (black arrow) likely due to the existence of an outer companion as uncovered by the frequency analysis (Sect.~\ref{sec:frequency_analysis}). To identify potential wide co-moving companion(s), we queried the \gaia\,DR3 catalog through CDS/Vizier within a radius of one degree around TOI-1130. We used a radius of 1 mas and 5 mas in parallax and in each proper motion direction (right ascension and declination), respectively, to account for potential differences at large distances. This search did not return any companion. Following the procedure described in \citetads{2017MNRAS.464.2708S}, we constrained the properties of the outer planet \textbf{ $M_\mathrm{d}\, {a_\mathrm{d}}^{-2}\,=\,2.62\,\mjup\,\mathrm{AU}^{-2}$}. If we assume a circular orbit for the outer companion and take our 70-day RV baseline, we find an orbital period of $\sim$\,140 days corresponding to a semi-major axis of $a_\mathrm{d}\,=\,0.47$\,AU. This leads to a planet mass of \rev{$0.58$\,\mjup}. The maximal orbital distance assuming that the signal is planetary in nature ($M_\mathrm{d}\,=13$\,\mjup) is \rev{2.23\,AU}. The \gaia\,DR3 release indicates a low RUWE of 1.137 and a null value for the non-single star table entry. However, TOI-1130 has a large {\tt{sepsi}} parameter of 16.29, which indicates a significant excess in the astrometric noise \citepads{2018A&A...616A...2L}. This excess might be related to the long linear trend seen in the RV time series. 

The results derived here using the photodynamical model could change when considering a third planet in the system. However, since the third planet is to our best knowledge at the moment, further out, we expect no strong interactions with TOI-1130\,b and TOI-1130\,c and we can treat our result as valid. We find planetary masses for TOI-1130\,b and TOI-1130\,c of \rev{$M_\mathrm{b}\,=19.28\pm\,0.97$\,\mearth and $M_\mathrm{c}\,=325.69\pm\,5.59$\,\mearth, and radii of $R_\mathrm{b}\,=3.56\pm\,0.13$\,\rearth and $R_\mathrm{c}\,=13.32^{+1.55}_{-1.41}$\,\rearth, respectively}. TOI-1130\,c's planetary mass and orbital parameters were previously determined by \citetads{2020ApJ...892L...7H} based on RV measurements only. Our planetary mass and radius agree within 1\,$\sigma$ with their values\footnote{Converted from their original values of $M_\mathrm{c}=0.974^{+0.043}_{-0.044}$\,\mjup and $R_\mathrm{c}=1.50^{+0.27}_{-0.22}$\,\rjup.} of $M_\mathrm{c}=309.56^{+13.67}_{-13.98}$\,\mearth and $R_\mathrm{c}=~16.81^{+3.03}_{-2.47}$\,\rearth. We reach a precision of 2\% and \rev{11\%} in planet mass and radius, which translates into a precision of \rev{34\%($\rho_\mathrm{c}\,=0.75^{+0.31}_{-0.21}\mathrm{cm}^{-3}$)} in the mean density of TOI-1130\,c. We refine the orbital eccentricity to \rev{$e_\mathrm{c}\,=\,0.0457\pm0.0016$.} TOI-1130\,b was only validated in \citetads{2020ApJ...892L...7H}. We spectroscopically confirm the inner planet and measure its mass with a precision of \rev{5\%.} We find that TOI-1130\,b orbits the star on a slightly eccentric orbit with \rev{$e_\mathrm{b}\,=\,0.0541\pm0.0015$.} Our radius measurement of \rev{$R_\mathrm{b}\,=3.56\pm\,0.13$\,\rearth} agrees within 1\,$\sigma$ with $R_\mathrm{b}=3.65\pm0.10$\,\rearth determined by \citetads{2020ApJ...892L...7H}. We reach a precision of 4\% in the planet radius, which leads to a planetary density of \rev{$\rho_\mathrm{b}\,=\,2.34\pm\,0.26$\,g\,$\mathrm{cm}^{-3}$ with a precision of 11\%.} \rev{We find a small mutual inclination of $\mathrm{I}=1.11\pm0.39^\circ$} assuming $\Omega_\mathrm{b}=\pi$. 

    \label{fig:mass_eq}

\section{Discussion}
We use the new system parameters derived in the previous sections to make inferences about the system's dynamics and formations (Sect.~\ref{sec:dyna_prop}), as well as the planet's interiors (Sect.~\ref{sec:interior}). 

\subsection{Dynamics and formation}
\label{sec:dyna_prop}

\begin{figure*}[t]%
    \centering
    \includegraphics[width=\textwidth]{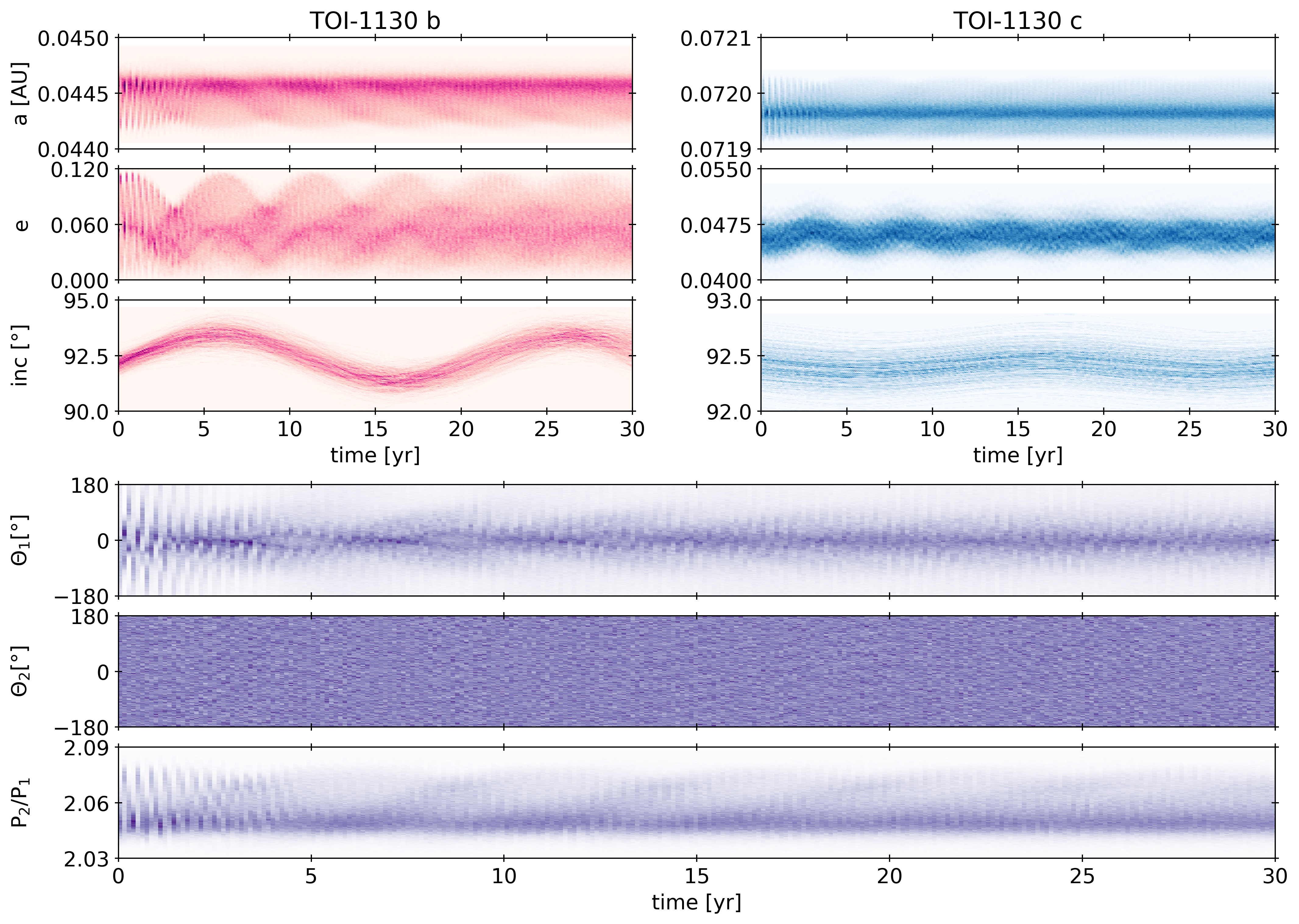} %
    \caption{Evolution of TOI-1130's orbital parameters reported in Table~\ref{tab:toi-1130_values} for 30 years. Top panels: Evolution of the planetary semi-major axes, orbital eccentricities, and inclinations of TOI-1130\,b (left) and TOI-1130\,c (right). Lower panels: Evolution of the 
    resonant \rev{angles $\Theta_\mathrm{1}$ and $\Theta_\mathrm{2}$}, and the period ratio of TOI-1130\,b and TOI-1130\,c. The color shade represents the probability of the \rev{quantity} having a certain value \rev{at a given time}, with a darker shade indicating a higher probability. \rev{These probabilities were calculated based on the 600 dynamical simulations discussed in the text.}
    }%
    \label{fig:oe}%
\end{figure*}

We performed a numerical stability analysis of the posterior from the photodynamical modeling to see if the planetary system lies in a stable configuration. The numerical simulations were carried out with the \rebound N-Body code with the IAS15 integrator. The system was simulated for $10^{7}$ years by drawing randomly 600 parameter combinations from the posterior in Table.~\ref{tab:toi-1130_values}. The posterior is found to lie in a stable configuration. 

\begin{figure*}[t]%
    \centering
    \includegraphics[width=\textwidth]{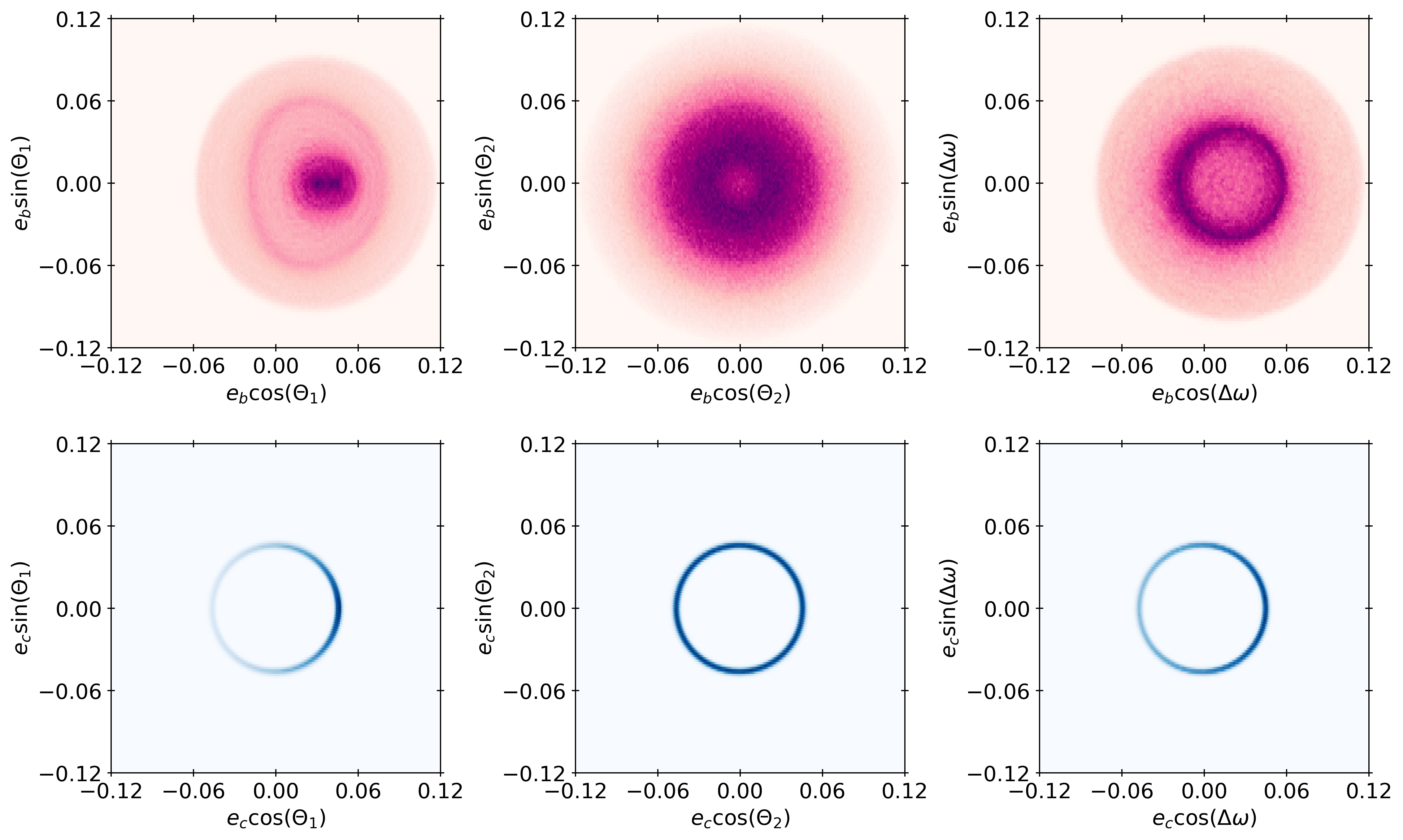} %
    \caption{Evolution of the trajectories of the secular apsidal angle and resonant angles. The color shade is the same as in Fig.~\ref{fig:oe}.
    }%
    \label{fig:angles}%
\end{figure*}

Since the system is close to a 2:1 period commensurability, we calculated the normalized distance to a $j:~j-1$ MMR as defined in \citetads{2012ApJ...761..122L}: 
\begin{equation}
    \Delta =\frac{P_ \mathrm{c}}{P_\mathrm{b}}\frac{j-1}{j}-1,
\end{equation}
where $P_ \mathrm{b}$ and $P_\mathrm{c}$ are the orbital periods of the inner and outer planets. Using \rev{our 600 samples, we find $\Delta\,=\,0.02554\,\pm\,0.00009$}, meaning that the two planets lie wide of their 2:1 resonance. To determine whether the planets are locked in resonance or not, we monitored the long-term evolution of the resonant angles during our stability simulation. For the 2:1 MMR, the resonant angles are defined as 
\begin{gather}
    \Theta_\mathrm{1} = 2\lambda_\mathrm{c} - \lambda_\mathrm{b} - \varpi_\mathrm{b}, \\
    \Theta_\mathrm{2} = 2\lambda_\mathrm{c} - \lambda_\mathrm{b} - \varpi_\mathrm{c},
\end{gather}
where $\lambda_\mathrm{i} = M_\mathrm{i} + \varpi_\mathrm{i}$ is the mean longitude, $M_\mathrm{i}$ is the mean anomaly and $\varpi_\mathrm{i} = \omega_\mathrm{i} + \Omega_\mathrm{i}$ is the longitude of periapse (see \citeads{1999ssd..book.....M}). 
Figure~\ref{fig:oe} shows the \rev{the results from the simulation: the} evolution of the semi-major axis, eccentricities, inclinations, 
the resonant \rev{angles} $\Theta_\mathrm{1}$ \rev{and $\Theta_\mathrm{2}$}, and the orbital period ratio of TOI-1130\,b and TOI-1130\,c for 30 years. The resonant angle $\Theta_\mathrm{1}$ shows a libration around $0^{\circ}$ \rev{with a period of \rev{$4.5\,\pm\,0.3\,$years}, while the amplitude is not constrained. The second resonant angle $\Theta_\mathrm{2}$ shows no libration and its amplitude is thus always $\pi$}. Figure~\ref{fig:angles} shows the evolution of the trajectories of the resonant angles $\Theta_1$ and $\Theta_2$. There is a libration visible supporting that the system is in a 2:1 MMR. \rev{A study by \citetads{2018AJ....155..106M} reports that the libration amplitudes of the resonant angles in the GJ 876 system seem to decrease with the amount of data used to characterize the system. Whether the libration amplitude of TOI-1130's resonant angle will become detectable when more data are used or not, will be seen in the future when more data (RV and photometry) become available.}

A planetary system in MMR needs \rev{either} to be assembled during the gas phase of the disk when the planets can undergo convergent migration and find their stable resonant configuration while their eccentricities and inclinations are damped by the gas disk \rev{or via modest tidal migration \citepads{2014A&A...566A.137D}}. In addition, a resonant system needs to survive the chaotic era after disk dispersal. 
Hence, multi-planet systems that are in MMR are well-preserved fossils from their formation era in the gas disk. Investigating their formation scenarios can provide clues to their natal disk properties (\citeads{2021A&A...656A.115H}; \citeads{2022MNRAS.511.3814H}). \rev{The gravitational interaction between two planets in MMR can excite the planet's eccentricity while gas disk-planet interaction damps them. The final eccentricities of planets' are determined by which MMR the planets are in and how much damping from the disk is applied on the planets (e.g. \citeads{2001A&A...374.1092S}; \citeads{2002ApJ...567..596L}; \citeads{2012ARA&A..50..211K})}. Because the outer planet TOI-1130\,c is much more massive than the inner one, it can give enough eccentricity to the inner planet to match our observations while it is difficult for TOI-1130\,c to become eccentric unless an additional mechanism interferes. \rev{For example comparing Fig. 4, and 5 of \citetads{2021A&A...648A..69A} clearly shows that when the mass of the outer planet doubles in a two-planet system, the eccentricity of the inner planet considerably increases while the eccentricity of the outer planet remains similar to the equal-mass model. In addition, in their Fig. 10, they presented the results of a model identical to Fig 4 but with an additional third planet. In this model, the eccentricity of the second planet increased greatly compared to the two-planet model as the result of gravitational interaction with the third planet. However, for the case of massive planets in MMR, the eccentricity damping/excitation could be complicated and needs a more detailed study (e.g. \citeads{2005A&A...437..727K}; \citeads{2018A&A...618A.169C}).} Another mechanism suggested by \citeads{2021MNRAS.500.1621D} needs the giant planet to be located at the appropriate position inside a disk inner cavity such that the Lindblad resonances which damp the planet's eccentricity reside inside the cavity while those that excite the planet's eccentricity lie in the disk, where the surface density is higher.

\rev{If the eccentricity of the hot Jupiter is a relic of formation, rather than ongoing planet--planet dynamics, it cannot have been damped by tidal interactions acting between the planet and the star over the system's lifetime. For systems such as this, eccentricity decay is driven mainly by the tidal distortion of the planet by the star. We estimate the decay timescale for a constant-$Q$ tidal model using equation~1 of \citepads{2008ApJ...678.1396J}:}
\begin{equation}
\frac{\dot{e}}{e} = -\frac{63}{4}\left(GM_\star^3\right)^{1/2} \frac{R_\mathrm{pl}^5}{Q_\mathrm{pl}M_\mathrm{pl}}a^{-13/2},
\end{equation}
\rev{where $Q_\mathrm{pl}$ is the tidal quality factor of the planet. We find a decay timescale of 5\,Gyr occurs for a quality factor $Q_\mathrm{pl}\approx5\times10^5$; for comparison, our Solar System's Jupiter has a quality factor around $10^5$ \citepads{2009Natur.459..957L} but a large spread of values is possible depending on the planet's internal structure and the frequencies of the tidal forcing. It is therefore not possible to rule out a primordial origin of the eccentricity in this system, but tidal forces may well be significant in the long-term orbital dynamics.}

\subsection{Interior modeling}
\label{sec:interior}
Knowing the radii and masses of several planets in the same system is extremely useful because one can remove the age uncertainty when comparing the planets to each other, thereby providing important constraints for formation models \citepads{2011A&A...531A...3H}. Here we use CEPAM \citepads{1995A&AS..109..109G} and a non-grey atmosphere \citepads{2015A&A...574A..35P} to model the evolution of both planets in the system. We assume simple structures consisting of a central dense core and a surrounding hydrogen and helium envelope of solar composition. The core is assumed to be made with 50\% of ices and 50\% of rocks.

\begin{figure}[!b]%
    \centering
    \includegraphics[width=\columnwidth]{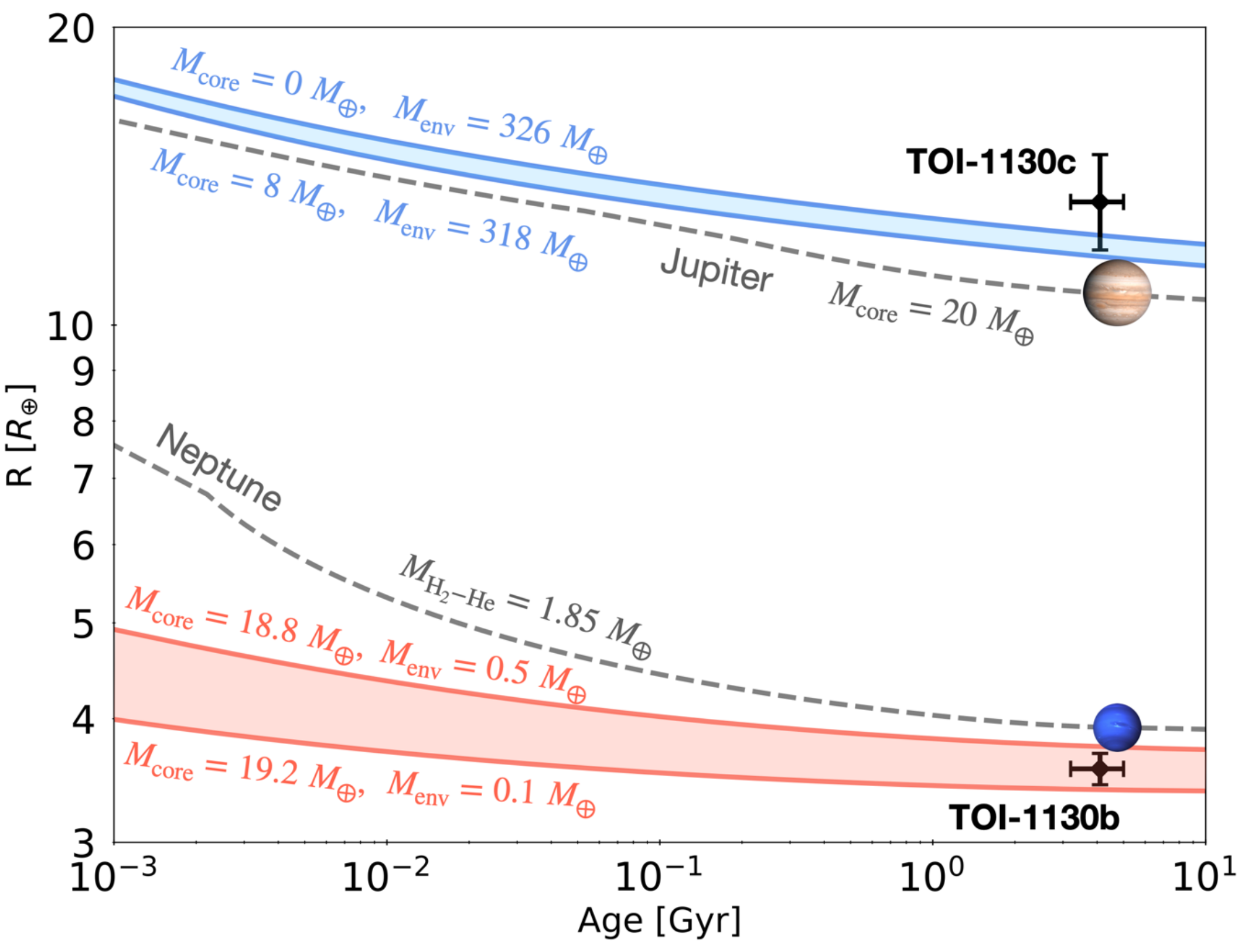} %
    \caption{Evolution models of TOI-1130\,b and TOI-1130\,c compared to Neptune and Jupiter, respectively. All models assume a central ice-rock core overlain by a solar composition hydrogen-helium envelope. \rev{$M_{\rm core}$ corresponds to the mass of the core while $M_{\rm env}$ corresponds to the mass of the envelope.} The range of envelope mass compatible with the observational constraints is shown for TOI-1130\,b and compared to a similarly simple model of Neptune. For TOI-1130\,c, only an upper limit on the core mass can be derived. For all cases, additional uncertainties on the core and envelope masses arise due to the unknown interior composition, temperature structure, and equation of state uncertainties. The black error bars correspond to observational constraints on the age and the radius of TOI-1130\,b and TOI-1130\,c.}%
    \label{fig:interior}%
\end{figure}

Figure~\ref{fig:interior} shows the resulting evolution models and observational constraints for both planets in the system. For guidance, we compare them to similarly simple models of Jupiter and Neptune (which have very similar masses as TOI-1130\,c and TOI-1130\,b, respectively), knowing that the ensemble of possibilities regarding their structure and composition is much wider \citepads{2020RSPTA.37890474H}. TOI-1130\,c is found to have a small enrichment in heavy elements, with a core of less than \rev{8}\,\mearth. This is \rev{lower} to what was obtained for our simple model of Jupiter. We point out however that \rev{Jupiter's enrichment} corresponds to a much wider range of possibilities, i.e. 8 - 46\,\mearth \citepads{2022arXiv220504100G}. TOI-1130\,b
is slightly more massive than Neptune but is found to contain less hydrogen and helium, assuming a similar composition core. With the same hypotheses, as shown in Fig.~\ref{fig:interior}, the envelope of TOI-1130\,b must be smaller than 0.5\,\mearth (i.e., less than 3\% of the mass of TOI-1130\,b compared to about 10\% of the mass of Neptune). 

Determining atmospheric abundances and possibly ice-to-rock ratios in TOI-1130\,b and TOI-1130\,c will be key to understanding the structure of ice and gas giants and the formation of these planetary systems \citepads{2022arXiv220504100G}.

\section{Conclusions}

We present a photodynamical analysis of the TOI-1130 planetary system that models consistently the \tess photometry, the HARPS and PFS RV measurements, and the ground-based photometry, accounting for the gravitational interaction between the bodies. The outer planet, TOI-1130\,c, is a hot Jupiter that was previously detected and confirmed in \citetads{2020ApJ...892L...7H}, while the inner Neptune-sized planet, TOI-1130\,b, was only validated. Here, both planets are confirmed and precisely characterized in terms of orbital parameters and planetary masses \rev{($M_\mathrm{b}\,=19.28\pm\,0.97$\,\mearth and $M_\mathrm{c}\,=325.69\pm\,5.59$\,\mearth)} down to a precision of \rev{5}\,\% and 2\,\%, respectively. Due to the high impact parameter of the transit of TOI-1130\,c, we can only determine its radius with a precision of \rev{34}\,\% ($R_\mathrm{c}\,=13.32^{+1.80}_{-1.21}$\,\rearth) and its density with a precision of 35\,\% \rev{($\rho_\mathrm{c}\,=0.75^{+0.31}_{-0.21}$\,g\,$\mathrm{cm}^{-3}$)}. We find the radius of TOI-1130\,b to be \rev{$R_\mathrm{b}\,=3.56\pm\,0.13$\,\rearth} (4\%), which translates into a precision of 11\,\% for its mean density \rev{($\rho_\mathrm{b}\,=\,2.34\pm\,0.26$\,g\,$\mathrm{cm}^{-3}$)}. The RV follow-up observations we carried out with the HARPS and PFS spectrographs unveiled a significant acceleration very likely induced by a massive outer companion.

Our mass estimates are mainly driven by the RVs and less by the photometry (Fig.~\ref{fig:ttv_rv_contour}). TOI-1130 should be monitored to allow for an independent mass measurement via photometry alone. \tess will re-observe TOI-1130 in Sector~67, until then further transit follow-up observations from  the ground or from space, e.g., with the CHaracterising ExOPlanets Satellite (\textit{CHEOPS}), are recommended to get an improved phase coverage of the TTVs. \rev{Furthermore, additional observations can confirm or reject the presence of the potential third planet.}

TOI-1130 joins the small sub-sample of hot Jupiters with a nearby inner companion \rev{like WASP-47, Kepler-730, WASP-132, and TOI-2000}. Our results show that TOI-1130\,b and TOI-1130\,c are most likely in 2:1 MMR \rev{which puts TOI-1130 in a unique position in this sample. Although it is not the first giant planet system found to be in a 2:1 resonant configuration, it is the first known system with a close-in gas giant and an inner low-mass planet locked in a 2:1 MMR. Other giant planet systems which are known to lie in such a resonant configuration consist of only warm giant planets 
( e.g., HD 82943 \citepads{2013ApJ...777..101T}, TOI-216 (\citeads{2019MNRAS.486.4980K}: \citeads{2019AJ....158...65D}, \citeyearads{2021AJ....161..161D}), TIC 279401253 \citepads{2023ApJ...946L..36B}). However, several giant planet systems with and without low-mass planets are known to lie near but not in a 2:1 resonance (e.g., Kepler-30 (\citeads{2012ApJ...750..114F}; \citeads{2018MNRAS.478.2480P}; \citeads{2018AJ....156...96W}, \citeads{2022AJ....164...42J}), Kepler-56 (\citeads{2013MNRAS.428.1077S}; \citeads{2013Sci...342..331H}; \citeads{2016AJ....152..165O}), Kepler-88 (\citeads{2013ApJ...777....3N}; \citeads{2014A&A...561L...1B}), Kepler-89 (\citeads{2013ApJ...778..185M}; \citeads{2013ApJ...768...14W}; \citeads{2021MNRAS.503.4092B}; \citeads{2022AJ....164...42J}), TOI-2202 \citepads{2021AJ....162..283T}, TOI-2525 \citepads{2023AJ....165..179T}).}

\rev{Discoveries like TOI-1130 provide an important window into planetary formation.} Such a system contradicts the formation pathway for hot Jupiters via HEM, which prohibits the formation of planets inside the orbit of the hot Jupiter. Formation via disk migration or \rev{in situ} instead allows the existence of low-mass planets inside the orbit of a hot Jupiter. While planets that formed \rev{in situ} show a wide range of period ratios, planets on near-resonant orbits seem to be more likely an outcome of disk migration \rev{or modest tidal migration}. Therefore, it is more probable that this system is formed via migration than \rev{in situ}. We note that even if the MMR favors the formation via disk migration, the observed low eccentricities are also consistent with the \rev{in situ} formation. Therefore, further monitoring of this system is needed to distinguish between the two scenarios. 

Atmospheric characterization of this system with the James Webb Space Telescope could help to distinguish between the formation \rev{models}. Specifically, measuring the C/O abundances ratio could reveal whether the planets formed \rev{in situ} or beyond the snow line \rev{(\citeads{2011ApJ...743L..16O}; \citeads{2014ApJ...794L..12M}; \citeyearads{2017MNRAS.469.4102M}; \citeads{2017MNRAS.469.3994B})}. \rev{We expect the planets to have different atmospheric compositions if they formed far away from each other, while the compositions should be similar if they formed close to each other. Furthermore, the C/O ratio gives us clues about where in the disk the planets have formed. However, measuring the C/O ratio is challenging since it requires precise measurements of C and O-bearing molecules which account for around 70\% of the absorbers such as H$_2$O, CO, CO$_2$, and CH$_4$ and a good understanding of the underlying chemistry (\citeads{2021Natur.598..580L}; \citeads{2023MNRAS.520.4683F}; \citeads{2023MNRAS.tmp.1160B}).}

\begin{acknowledgements}
This work was supported by the KESPRINT collaboration, an international consortium devoted to the characterization and research of exoplanets discovered with space-based missions (\url{https://kesprint.science/}).  Based on observations carried out with the HARPS spectrograph mounted at the ESO 3.6-m telescope at La Silla Observatory under programme IDs 1102.C-0923 and 60.A-9709. This work makes use of observations from the Las Cumbres Observatory global telescope network. \rev{This work makes use of observations from the ASTEP telescope. ASTEP benefited from the support of the French and Italian polar agencies IPEV and PNRA in the framework of the Concordia station program, from Idex UCAJEDI (ANR-15-IDEX-01), OCA, ESA and the University of Birmingham.} This paper includes data collected with the \tess mission, obtained from the MAST data archive at the Space Telescope Science Institute (STScI). Funding for the \tess mission is provided by the NASA's Science Mission Directorate. STScI is operated by the Association of Universities for Research in Astronomy, Inc., under NASA contract NAS 5–26555. This research has made use of the Exoplanet Follow-up Observation Program website, which is operated by the California Institute of Technology, under contract with the National Aeronautics and Space Administration under the Exoplanet Exploration Program. We acknowledge the use of public TESS Alert data from pipelines at the TESS Science Office and at the TESS Science Processing Operations Center. Resources supporting this work were provided by the NASA High-End Computing (HEC) Program through the NASA Advanced Supercomputing (NAS) Division at Ames Research Center for the production of the SPOC data products. This work has made use of data from the European Space Agency (ESA) mission \gaia (\url{https://www.cosmos.esa.int/gaia}), processed by the \gaia Data Processing and Analysis Consortium (DPAC, \url{https://www.cosmos.esa.int/web/gaia/dpac/consortium}). Funding for the DPAC has been provided by national institutions, in particular the institutions participating in the \gaia Multilateral Agreement. This research has made use of the VizieR catalogue access tool, CDS, Strasbourg, France. This publication makes use of data products from the Two Micron All Sky Survey, which is a joint project of the University of Massachusetts and the Infrared Processing and Analysis Center/California Institute of Technology, funded by the National Aeronautics and Space Administration and the National Science Foundation. This publication makes use of data products from the Wide-field Infrared Survey Explorer, which is a joint project of the University of California, Los Angeles, and the Jet Propulsion Laboratory/California Institute of Technology, funded by the National Aeronautics and Space Administration. We are extremely grateful to the ESO staff members for their unique and superb support during the observations, and to François Bouchy and Xavier Dumusque for coordinating the HARPS time sharing agreement. \rev{J.~K. gratefully acknowledges the support of the Swedish National Space Agency (SNSA; DNR 2020-00104) and of the Swedish Research Council  (VR: Etableringsbidrag 2017-04945)}. A.~M. and C.~M.~P. gratefully acknowledge the support of the Swedish National Space Agency (SNSA; DNR 2020-00104, 177/19, 174/18, and 65/19). M.~E. acknowledges the support of the DFG priority program SPP 1992 “Exploring the Diversity of Extrasolar Planets” (HA 3279/12-1). J.~S. and P.~K. acknowledge support from the MSMT grant LTT-20015. J.~S. would like to acknowledge support from the Grant Agency of Charles University: GAUK No. 314421. R.~L. acknowledges funding from University of La Laguna through the Margarita Salas Fellowship from the Spanish Ministry of Universities ref. UNI/551/2021-May 26, and under the EU Next Generation funds. S.~M. acknowledges support by the Spanish Ministry of Science and Innovation with the Ramon y Cajal fellowship number RYC-2015-17697 and the grant number PID2019-107187GB-I00. N.~L. acknowledges support from the Agencia  Estatal  de  Investigaci\'on  del  Ministerio  de  Ciencia  e Innovaci\'on (AEI-MCINN) under grant PID2019-109522GB-C53\@. This work is partly supported by JSPS KAKENHI Grant Numbers JP19K14783 and JP21H00035. \rev{A.~J.~M. acknowledges support from the Swedish National Space Agency (grant 120/19C) and the Swedish Research Council (grant 2017-04945).}
\end{acknowledgements}
%
%
\bibliographystyle{aa} 
\bibliography{bibliography.bib} 

\begin{appendix} 
\section{Radial velocities}
\begin{table*}
    \centering
    \footnotesize
    \caption{Radial velocity measurements and spectroscopic activity indicators for TOI-1130 from HARPS spectra}
    \label{tab:toi-1130_harps}
    \begin{tabularx}{\textwidth}{@{\extracolsep{\fill}}ccccccccc}
    \toprule
    \toprule
    BJD$_\mathrm{TDB}$ & RV & $\sigma$RV & BIS & FWHM & S-index & $\sigma$S-index & S/N & $t_\mathrm{exp}$ \\
    -2457000 & [m\,$\mathrm{s}^{-1}$] & [m\,$\mathrm{s}^{-1}$] & [m\,$\mathrm{s}^{-1}$] & [m\,$\mathrm{s}^{-1}$] & - & - & @ 550 nm & [s] \\
    \midrule
    1744.516586 & -8071.84 & 7.19 & 53.17 & 6394.96 & 0.799 & 0.133 & 18.0 & 2100 \\
    1745.491812 & -7981.02 & 2.15 & 51.91 & 6307.82 & 0.699 & 0.025 & 43.7 & 2100 \\
    1746.506986 & -7887.07 & 1.76 & 52.82 & 6313.49 & 0.749 & 0.020 & 51.6 & 2100 \\
    1747.534521 & -7847.40 & 1.84 & 50.74 & 6308.11 & 0.732 & 0.023 & 50.1 & 2100 \\
    1748.495349 & -7884.41 & 1.86 & 47.21 & 6296.11 & 0.741 & 0.020 & 49.0 & 2100 \\
    1752.548748 & -8091.47 & 1.88 & 49.45 & 6313.41 & 0.762 & 0.018 & 45.5 & 2100 \\
    1753.514551 & -8009.94 & 1.46 & 46.58 & 6314.93 & 0.743 & 0.014 & 57.5 & 2100 \\
    1754.513913 & -7911.19 & 1.69 & 53.57 & 6317.25 & 0.750 & 0.016 & 50.0 & 2100 \\
    1755.513265 & -7849.18 & 1.45 & 53.29 & 6323.09 & 0.736 & 0.013 & 57.6 & 2100 \\
    1755.596208 & -7845.92 & 1.55 & 43.68 & 6318.08 & 0.747 & 0.019 & 55.8 & 2100 \\
    1756.508322 & -7864.35 & 2.02 & 49.33 & 6310.58 & 0.745 & 0.022 & 43.3 & 2100 \\
    1756.574080 & -7871.85 & 2.31 & 41.50 & 6324.93 & 0.759 & 0.029 & 39.2 & 2100 \\
    1757.502187 & -7936.45 & 2.22 & 44.92 & 6313.49 & 0.800 & 0.025 & 40.1 & 2100 \\
    1757.581857 & -7941.94 & 2.35 & 51.72 & 6319.50 & 0.806 & 0.029 & 38.6 & 2100 \\
    1759.567931 & -8082.26 & 3.49 & 54.33 & 6322.03 & 0.853 & 0.043 & 28.1 & 2100 \\
    1761.551634 & -8032.14 & 1.77 & 36.48 & 6323.26 & 0.754 & 0.022 & 49.4 & 2100 \\
    1761.601212 & -8026.75 & 1.73 & 47.81 & 6329.05 & 0.815 & 0.023 & 51.1 & 2100 \\
    1762.558170 & -7924.98 & 1.56 & 49.49 & 6326.84 & 0.729 & 0.017 & 55.1 & 2100 \\
    1762.622204 & -7923.65 & 1.94 & 45.11 & 6331.03 & 0.793 & 0.025 & 46.2 & 2100 \\
    1763.534758 & -7849.78 & 1.50 & 52.98 & 6323.12 & 0.793 & 0.015 & 56.8 & 2100 \\
    1764.563457 & -7848.33 & 1.96 & 46.44 & 6321.45 & 0.779 & 0.021 & 44.8 & 2100 \\
    1766.527451 & -7980.54 & 1.35 & 55.60 & 6332.03 & 0.805 & 0.015 & 63.6 & 2100 \\
    1766.581646 & -7989.46 & 1.47 & 53.24 & 6332.32 & 0.773 & 0.017 & 59.2 & 2100 \\
    1768.521916 & -8098.51 & 1.44 & 48.83 & 6319.00 & 0.792 & 0.015 & 59.5 & 2100 \\
    1768.547573 & -8099.68 & 1.50 & 47.68 & 6323.98 & 0.776 & 0.016 & 57.3 & 2100 \\
    1773.512029 & -7875.94 & 3.65 & 34.78 & 6327.09 & 0.711 & 0.054 & 29.6 & 2100 \\
    1774.564408 & -7954.29 & 2.92 & 57.44 & 6327.46 & 0.764 & 0.042 & 35.2 & 2100 \\
    1775.507737 & -8026.29 & 2.43 & 44.21 & 6316.48 & 0.772 & 0.032 & 40.3 & 2100 \\
    1777.535606 & -8084.98 & 4.77 & 60.32 & 6300.79 & 0.757 & 0.084 & 24.3 & 2100 \\
    1781.551841 & -7860.85 & 1.54 & 49.56 & 6313.50 & 0.749 & 0.019 & 56.2 & 2100 \\
    1782.543742 & -7922.02 & 1.91 & 50.68 & 6310.09 & 0.797 & 0.021 & 45.8 & 2100 \\
    1783.520045 & -7998.14 & 1.44 & 50.81 & 6322.14 & 0.897 & 0.013 & 58.1 & 2100 \\
    1784.520593 & -8070.87 & 2.35 & 54.33 & 6312.80 & 0.804 & 0.029 & 38.7 & 2100 \\
    1785.515228 & -8089.91 & 2.58 & 45.32 & 6327.29 & 0.812 & 0.029 & 35.5 & 2100 \\
    1791.499671 & -7962.54 & 4.61 & 76.81 & 6267.20 & 0.645 & 0.039 & 23.9 & 2100 \\
    1792.517829 & -8043.12 & 1.63 & 54.64 & 6336.37 & 0.859 & 0.019 & 53.6 & 2100 \\
    1793.533452 & -8097.38 & 2.04 & 61.37 & 6336.75 & 0.826 & 0.026 & 44.0 & 2100 \\
    1794.524991 & -8052.72 & 1.69 & 55.51 & 6332.86 & 0.805 & 0.019 & 51.7 & 2100 \\
    1794.568008 & -8054.37 & 2.26 & 48.31 & 6334.79 & 0.774 & 0.029 & 40.8 & 2100 \\
    1795.526473 & -7949.96 & 2.15 & 46.31 & 6318.66 & 0.815 & 0.022 & 41.2 & 2100 \\
    1796.518281 & -7861.47 & 1.52 & 49.39 & 6318.77 & 0.792 & 0.018 & 57.3 & 2100 \\
    1797.519454 & -7832.17 & 2.03 & 54.43 & 6316.68 & 0.749 & 0.022 & 43.8 & 2100 \\
    1798.538403 & -7864.31 & 2.22 & 54.98 & 6311.97 & 0.657 & 0.026 & 43.2 & 2100 \\
    1799.528154 & -7934.69 & 2.44 & 48.59 & 6291.12 & 0.775 & 0.027 & 37.8 & 2100 \\
    1802.518128 & -8081.30 & 1.68 & 41.05 & 6319.98 & 0.751 & 0.020 & 52.1 & 2100 \\
    1803.515728 & -7987.21 & 1.69 & 54.63 & 6318.04 & 0.794 & 0.024 & 52.4 & 2100 \\
    1804.527948 & -7879.48 & 1.96 & 48.16 & 6309.01 & 0.834 & 0.028 & 45.7 & 2100 \\
    1805.517277 & -7831.15 & 2.36 & 45.07 & 6307.73 & 0.766 & 0.030 & 38.9 & 2100 \\
    1806.515287 & -7846.64 & 2.09 & 40.90 & 6302.85 & 0.718 & 0.028 & 43.3 & 2100 \\
        \bottomrule
        \bottomrule
            \end{tabularx}
\end{table*}

\begin{table*}
    \centering
    \small
    \caption{Radial velocity measurements and spectroscopic activity indicators for TOI-1130 from PFS spectra. 5\% fractional uncertainty is assumed for the S-index and the H$\alpha$-index.}
    \label{tab:toi-1130_pfs}
    \begin{tabularx}{\textwidth}{@{\extracolsep{\fill}}cccccccc}
    \toprule
    \toprule
    BJD$_\mathrm{TDB}$ & RV & $\sigma$RV & S-index & $\sigma$S-index & H$\alpha$-index & $\sigma$H$\alpha$-index & $t_\mathrm{exp}$ \\
    -2457000 & [m\,$\mathrm{s}^{-1}$] & [m\,$\mathrm{s}^{-1}$] & - & - & - & - & [s] \\
    \midrule
    1738.530612 & 195.71 & 1.40 & 0.766 & 0.038 & 0.0478 & 0.0024 & 1200 \\
    1738.545340 & 196.86 & 1.51 & 0.870 & 0.043 &  0.0479 & 0.0024 & 1200 \\
    1739.523756 & 207.47 & 1.14 & 0.775 & 0.039 & 0.0475 & 0.0034 & 1200 \\
    1739.537905 & 206.97 & 1.27 & 0.824 & 0.041 & 0.0473 & 0.0024 & 1200 \\
    1741.595996 & 46.13 & 1.16 & 0.803 & 0.040 & 0.0470 & 0.0024 & 1200 \\
    1741.612304 & 46.22 & 1.16 & 0.778 & 0.039 & 0.0476 & 0.0024 & 1401 \\
    1761.520197 & 29.52 & 1.10 & 0.727 & 0.036 & 0.0476 & 0.0024 & 1200 \\
    1761.534865 & 27.71 & 1.37 & 0.804 & 0.040 & 0.0455 & 0.0023 & 1200 \\
    1767.524532 & -0.34 & 1.33 & 0.829 & 0.041 & 0.0478 & 0.0024 & 1200 \\
    1767.539261 & 0.00 & 1.14 & 0.808 & 0.040 & 0.0475 & 0.0024 & 1200 \\
    1767.553969 & 1.45 & 1.57 & 0.898 & 0.045 & 0.0474 & 0.0024 & 1200 \\
    1768.520979 & -35.39 & 1.43 & 0.856 & 0.043 & 0.0455 & 0.0023 & 1200 \\
    1768.534587 & -33.59 & 1.06 & 0.839 & 0.042 & 0.0475 & 0.0024 & 1200 \\
    1768.549296 & -34.44 & 1.16 & 0.846 & 0.042 & 0.0475 & 0.0024 & 1200 \\
    1768.563915 & -39.56 & 1.15 & 0.808 & 0.040 & 0.0480 & 0.0024 & 1200 \\
    1768.577673 & -38.01 & 1.40 & 0.869 & 0.043 & 0.0478 & 0.0024 & 1200 \\
    1768.608230 & -32.8 & 1.71 & 0.903 & 0.045 & 0.0479 & 0.0024 & 1200 \\
    1768.623109 & -42.27 & 1.84 & 0.873 & 0.044 & 0.0480 & 0.0024 & 1200 \\
    1768.638098 & -34.79 & 1.88 & 0.907 & 0.045 & 0.0468 & 0.0023 & 1200 \\
    1768.652626 & -34.61 & 1.65 & 0.876 & 0.044 & 0.0475 & 0.0024 & 1200 \\
        \bottomrule
        \bottomrule
            \end{tabularx}
\end{table*}

\section{Additional figures}

\begin{figure*}%
    \centering
    \includegraphics[width=\textwidth]{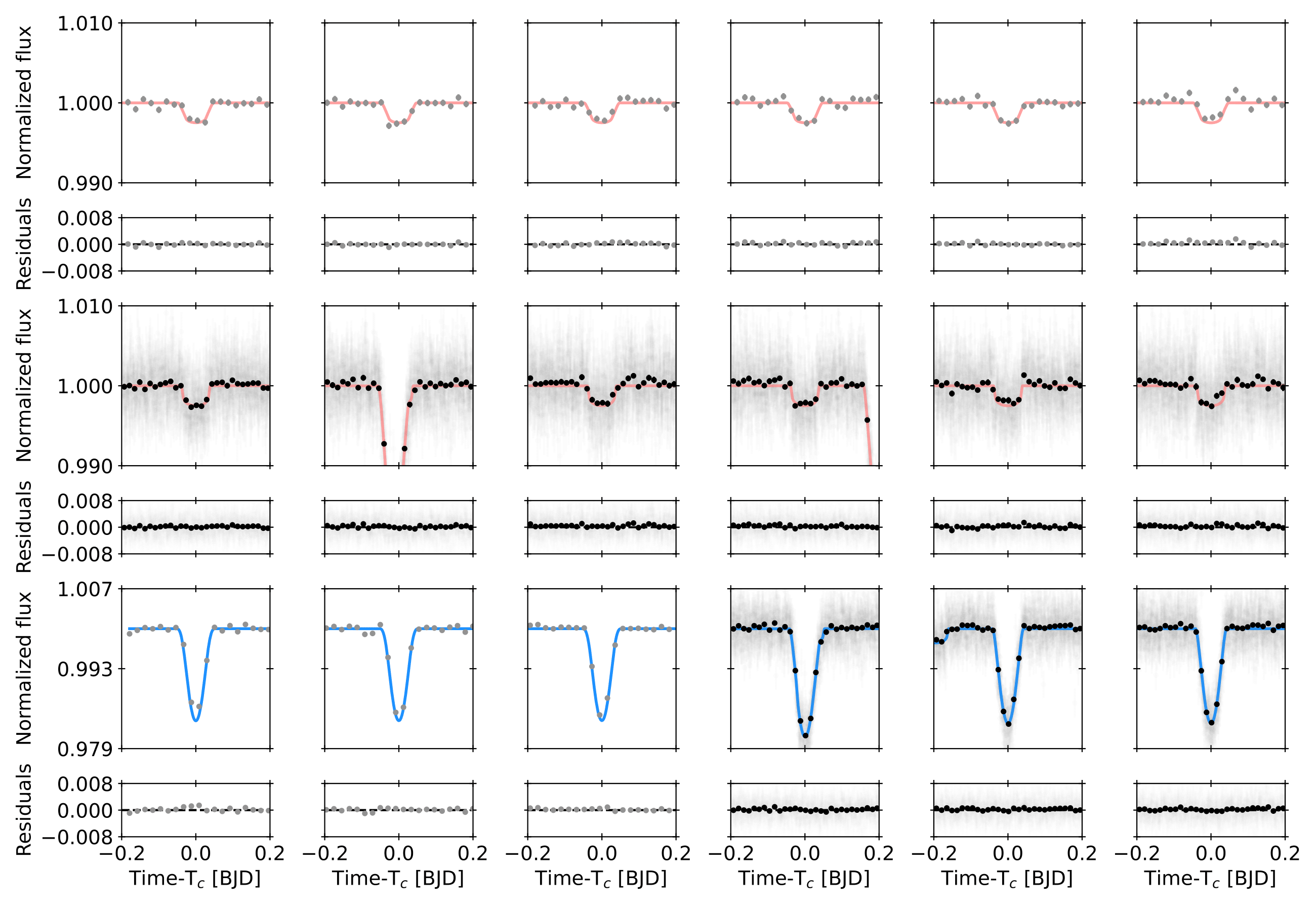}%
    \caption{\tess transits of TOI-1130\,b (upper two panels) and TOI-1130\,c (lower panel) as used in the photodynamical modeling. Each transit is centred around the individual transit center times. The median \pyttv photodynamical model (see Sect.~\ref{sec:photo_model}) is overplotted for both TOI-1130\,b and TOI-1130\,c color-coded in salmon and blue, respectively. The light curves from Sector~27 are binned to 20-min (black points). Residuals are shown beneath each plot.}%
    \label{fig:transits_photodyn_b}%
\end{figure*}
\end{appendix}
\end{document}